\RequirePackage[l2tabu,orthodox]{nag}
\documentclass[aps,prd,twocolumn,superscriptaddress,showkeys,nofootinbib,preprintnumbers]{revtex4-1}

\usepackage[T1]{fontenc}
\usepackage[utf8]{inputenc}
\usepackage{latexsym}
\usepackage{amsmath}
\usepackage{amssymb}
\usepackage{amsfonts}
\usepackage{color}
\usepackage{tikz}
\usepackage{supertabular} 
\usepackage{placeins}
\usepackage{graphicx}
\usepackage{dsfont}
\usepackage{capt-of}
\usepackage{hyperref}
\usepackage{bm}

\renewcommand{\O}{\mathcal{O}}
\renewcommand{\l}{\ell}
\renewcommand{\d}{\mathrm{d}}

\newcommand{\e}[1]{\mathrm{e}^{#1}}
\renewcommand{\ln}[1]{\mathrm{ln}\!\left({#1}\right)}
\newcommand{\lnSquare}[1]{\mathrm{ln}^2\!\left({#1}\right)}

\newcommand{\lnEne}[1]{\mathrm{ln}^n\!\left({#1}\right)}

\renewcommand{\r}[2]{\rho_{#1,#2}}
\newcommand{\M}{\mathcal{M}}

\begin{document}

\preprint{TUM-EFT 86/16}

\title{Electric dipole transitions of $1P$ bottomonia}

\author{Jorge Segovia}
\affiliation{Departamento de Sistemas F\'isicos, Qu\'imicos y Naturales,
Universidad Pablo de Olavide, 41013 Sevilla, Spain}

\author{Sebastian Steinbei{\ss}er}
\affiliation{Physik-Department, Technische Universit\"at M\"unchen, 
James-Franck-Stra{\ss}e 1, 85748 Garching, Germany}

\author{Antonio Vairo}
\affiliation{Physik-Department, Technische Universit\"at M\"unchen, 
James-Franck-Stra{\ss}e 1, 85748 Garching, Germany}

\date{\today}

\begin{abstract}
We compute the electric dipole transitions $\chi_{bJ}(1P) \to \gamma 
\Upsilon(1S)$, with $J=0,1,2$, and $h_{b}(1P)\to \gamma\eta_{b}(1S)$ in a 
model-independent way. We use potential non-relativistic QCD (pNRQCD) at weak 
coupling with either the Coulomb potential or the complete static potential 
incorporated in the leading order Hamiltonian. In the last case, the 
perturbative series shows very mild scale dependence and a good convergence 
pattern, allowing predictions for all the transition widths. Assuming 
$\Lambda_{\text{QCD}} \ll mv^2$, the precision that we reach is 
$k_{\gamma}^{3}/(mv)^{2} \times \mathcal{O}(v^{2})$, where $k_{\gamma}$ is the 
photon energy, $m$ is the mass of the heavy quark and $v$ its relative 
velocity. Our results are: $\Gamma(\chi_{b0}(1P)\to \gamma\Upsilon(1S)) = 
28^{+2}_{-2}~\text{keV}$, $\Gamma(\chi_{b1}(1P)\to \gamma\Upsilon(1S)) = 
37^{+2}_{-2}~\text{keV}$, $\Gamma(\chi_{b2}(1P)\to \gamma\Upsilon(1S)) = 
45^{+3}_{-3}~\text{keV}$ and $\Gamma(h_b(1P)\to \gamma\eta_b(1S)) = 
63^{+6}_{-6}~\text{keV}$.
\end{abstract}


\maketitle

\section{Introduction}
\label{sec:introduction}
Electromagnetic transitions are often a significant decay mode for bottomonium 
states below the $B\bar{B}$ threshold ($10.56\,\text{GeV}$), making them a 
suitable experimental tool to access lower states. For instance, the first 
$b\bar{b}$ states not directly produced in $e^{+}e^{-}$ collisions were the six 
triplet-$P$ states, $\chi_{bJ}(2P)$ and $\chi_{bJ}(1P)$, with $J=0,\,1,\,2$, 
discovered in radiative decays of the $\Upsilon(3S)$ and $\Upsilon(2S)$ in 
$1982$~\cite{Han:1982zk,Eigen:1982zm} and 
$1983$~\cite{Klopfenstein:1983nx,Pauss:1983pa}, respectively.

Electromagnetic transitions can be classified in terms of electric and magnetic 
multipoles. The most important ones are the E1 (electric dipole) and the M1 
(magnetic dipole) transitions; higher order multipole modes E2, M2, E3, 
$\ldots$ appear in the spectrum, but are suppressed. The width of allowed 
(hindered) M1 transitions is of order $k_{\gamma}^{3}/m^2$ 
($k_{\gamma}^{3}v^2/m^2$) where $k_{\gamma}$ is the photon energy and $m$ is 
the mass of the heavy quark, whereas the width of E1 transitions is of order 
$k_{\gamma}^{3}/(mv)^2$, where $v$, which is much smaller than 1, is the 
relative velocity of the heavy quarks in the 
quarkonium~\cite{Brambilla:2004wf}. Electric dipole transitions happen much 
more frequently than magnetic dipole transitions. The branching fraction for E1 
transitions is indeed significant for the bottomonium states that we shall 
study in this work~\cite{PhysRevD.98.030001}: $\mathcal{B}(\chi_{b0}(1P)\to 
\gamma\Upsilon(1S)) = (1.94 \pm 0.27)\,\%$, $\mathcal{B}(\chi_{b1}(1P)\to 
\gamma\Upsilon(1S)) = (35.0 \pm 2.1)\,\%$, $\mathcal{B}(\chi_{b2}(1P)\to 
\gamma\Upsilon(1S)) = (18.8 \pm 1.1)\,\%$, and $\mathcal{B}(h_{b}(1P)\to 
\gamma\eta_{b}(1S)) = (52^{+6}_{-5})\%$. Even in the $\chi_{b0}$ case this is 
the largest observed exclusive branching fraction.

Electric dipole transitions are characterized by the fact that they change 
the orbital angular momentum of the state by one unit, but not the spin. 
Therefore, the final state has different parity and C-parity than the initial 
one. Typical examples of E1 quarkonium decays are the ones mentioned above: 
$2^{3}P_{J}\to 1^{3}S_{1}+\gamma$ and $2^{1}P_{1}\to 1^{1}S_{0}+\gamma$. Here 
and in the following we denote the states as $n\,^{2s+1}\!\ell_{J}$, where 
$n=n_{r}+\ell+1$ is the principal quantum number, with $n_{r}=0,\,1,\,\ldots$ 
the radial quantum number and $\ell$ the orbital angular momentum usually 
represented by a letter: $S$ for $\ell=0$, $P$ for $\ell=1$ and so on. The spin 
is denoted by $s$ and $J$ is the total angular momentum. We use also the PDG 
notation, where $\chi_{bJ}(1P)$ identifies the state $2\,^{3}\!P_{J}$, and 
$h_{b}(1P)$ the state $2\,^{1}\!P_{1}$. This is to say, in the PDG notation, 
$1P$ bottomonia are states with quantum numbers $n=2$ and $\ell=1$.

E1 (and M1) electromagnetic transitions between heavy quarkonia have been 
treated for a long time by means of potential models that use non-relativistic 
reductions of QCD-based quark--(anti-)quark interactions (see, for instance, 
Ref.~\cite{Segovia:2016xqb} for a recent application to the bottomonium 
system). However, the release in the last decade of a new large set of accurate 
experimental data, concerning electromagnetic reactions in the heavy quark 
sector, by B-factories (BaBar, Belle and CLEO), $\tau$-charm facilities 
(CLEO-c, BESIII) and even proton--(anti-)proton colliders (CDF, D0, LHCb, 
ATLAS, CMS)~\cite{Brambilla:2010cs, Brambilla:2014jmp} demands for systematic 
and model-independent treatments.

The aim of this paper is to compute the E1 transitions $\chi_{bJ}(1P)\to 
\gamma\Upsilon(1S)$, with $J=0,1,2$, and $h_{b}(1P)\to \gamma\eta_{b}(1S)$ 
using potential non-relativistic QCD (pNRQCD). Quarkonium is characterized by 
the hierarchy of energy scales:
\begin{equation}
m \gg p \sim mv  \gg E \sim mv^2\,,
\label{eq:hierarchy}
\end{equation}
where $p$ is the relative momentum of the heavy quarks, proportional to the 
inverse of the size of the quarkonium, and $E$ is the binding energy. The 
relative heavy quark velocity, $v$, is assumed to be $v \ll 1$, which qualifies 
quarkonium as a non-relativistic bound state. pNRQCD is a non-relativistic 
effective field theory that takes advantage of this hierarchy of scales by 
systematically computing quarkonium observables as expansions in 
$v$~\cite{Pineda:1997bj,Brambilla:1999xf} (see 
Refs.~\cite{Brambilla:2004jw,Pineda:2011dg} for reviews). In the case of 
radiative transitions another relevant scale is the photon energy, $k_\gamma$. 
The photon energy is about the energy gap between the initial and final 
quarkonium states: for allowed (hindered) M1 transitions it is of the order of 
$mv^4$ ($mv^2$), for E1 transitions it is of the order of $mv^2$. The theory 
for M1 transitions in pNRQCD has been developed in~\cite{Brambilla:2005zw} and 
extended to E1 transitions in~\cite{Brambilla:2012be}. 
Ref.~\cite{Brambilla:2012be} provides the theoretical basis for the present 
study, which aims at computing E1 transitions from $1P$ bottomonium states at 
relative order $v^2$, \textit{i.e.}, at order $k_\gamma^3/m^2$ in the 
transition width.

The specific details of the construction of pNRQCD depend on the relative size 
of the scale $mv^2$ with respect to $\Lambda_{\text{QCD}}$. In this paper, we 
assume that $m v^2\gg \Lambda_{\text{QCD}}$.\footnote{The following 
computations are valid also for  $m v^2\sim \Lambda_{\text{QCD}}$. What changes 
in this case is, however, the parametrical size of the non-perturbative 
corrections, see Sec.~\ref{subsec:RelativisticCorrectionsFock} and comments in 
the conclusion.} The propagation of a color singlet heavy quark-antiquark 
field, $\textrm{S}$, is described at relative order $v^2$ by the Lagrangian 
density:
\begin{equation}
\mathcal{L} = \int d^3 r \; \mathrm{Tr} \left\{ \mathrm{S}^{\dagger} 
\left( {i {\partial}_0 
+ \frac{{\vec{\nabla}}^2}{4m} 
+\frac{{\vec{\nabla}}_r^2}{m}
+ \frac{{\vec{\nabla}}_r^4}{4m^3}} - V \right) \mathrm{S}\right\} \,,
\label{Lsinglet}
\end{equation}
where $r$ is the quark-antiquark distance parameterizing the color singlet 
field $\textrm{S}$ and $V$ is the quark-antiquark potential. The operator 
$-i{\vec{\nabla}} \sim mv^2$ is the center of mass momentum (the derivative 
acts on the center of mass coordinate), while  $-i{\vec{\nabla}}_r \sim mv$ is 
the relative momentum (the derivative acts on the distance $r$). If $m v^2\gg 
\Lambda_{\text{QCD}}$, the potential $V$ may be computed order by order in 
perturbation theory and $v\sim\alpha_{\textrm{s}}$, where $\alpha_{\textrm{s}}$ 
is the strong coupling evaluated at the typical momentum transfer scale. At 
leading order in $\alpha_{\textrm{s}}$, $V$ is given by the Coulomb potential 
between static color triplet and color antitriplet sources: 
$V_s^{(0)}=-4\alpha_{\textrm{s}}/(3r)$. According to the pNRQCD counting 
$V_s^{(0)} \sim mv^2$. E1 transitions are encoded in the part of the pNRQCD 
Lagrangian, $\mathcal{L}_{\gamma \textrm{pNRQCD}}$, that describes the 
interaction of the quark-antiquark field $\textrm{S}$ with the electromagnetic 
field:
\begin{equation}
\mathcal{L}_{\gamma \textrm{pNRQCD}} =
\int d^3 r \; \mathrm{Tr} \left\{ \mathrm{S}^{\dagger} {\vec{r}}\cdot e e_Q 
{\vec{E}}^{\textrm{em}} \mathrm{S}  + \dots \right\}\,.
\label{LagE1}
\end{equation} 
The displayed term is the leading order electric dipole interaction term 
($ee_Q$ stands for the electric charge of the heavy quark $Q$ and 
${\vec{E}}^{\textrm{em}}$ for the electric field), whereas the dots stand for 
higher order operators contributing to the E1 transition at relative order 
$v^2$ (or smaller), whose explicit expressions can be read off from 
Ref.~\cite{Brambilla:2012be}.

There seems to be a growing consensus in the literature that the weak-coupling 
regime $m v^2\gg \Lambda_{\text{QCD}}$ may indeed be applied to many physical 
observables in the bottomonium sector including $n=2$ bottomonium states
(for early work see~\cite{Brambilla:2001fw,Brambilla:2001qk,Brambilla:2004wu}, 
for reviews see~\cite{Brambilla:2010cs,Pineda:2011dg,Brambilla:2014jmp}, for 
recent work see~\cite{Sumino:2016sxe,Peset:2018jkf}). In order to reach this 
conclusion, it is crucial, however, to have a proper treatment for the large 
terms appearing in the perturbative expansion. As long as $\alpha_{\textrm{s}}$ 
remains a perturbative coupling, large terms can be due to factorially growing 
coefficients, which may require renormalon subtraction, or large logarithms in 
the renormalization scale.

In this work, we adopt methods to deal with both large corrections, eventually 
achieving a convergent expansion with mild dependence on the renormalization 
scale. Concerning the renormalon subtraction scheme, we adopt the one of 
Ref.~\cite{Pineda:2001zq}. Concerning the resummation of large logarithms, we 
rearrange the perturbative expansion of pNRQCD in such a way that the static 
potential is exactly included in the leading order (LO) Hamiltonian. This 
expansion scheme has been applied to the computation of the heavy quarkonium 
electromagnetic decay ratios in Ref.~\cite{Kiyo:2010jm} and to the 
determination of M1 transitions between low-lying heavy quarkonium states in 
Ref.~\cite{Pineda:2013lta}. The authors obtain agreement between theory and 
experiment for the case of the charmonium and bottomonium ground states and for 
the $n=2$ excitations of the bottomonium. Very recently, the same scheme has 
been applied to the spectrum of $n=2$, $l=1$ quarkonium states 
in~\cite{Peset:2018jkf}. Hence, another motivation for the present study is to 
probe weakly coupled pNRQCD in the context of electric dipole transitions from 
the spin-triplet and spin-singlet lowest bottomonium $P$-wave states.

In Ref.~\cite{Brambilla:2012be}, the complete set of relativistic corrections 
of relative order $v^2$ with respect to the leading order E1 decay width has 
been derived. In the E1 case, differently from M1 
transitions~\cite{Brambilla:2005zw,Pineda:2013lta}, the computation of 
relativistic corrections at relative order $v^2$ is technically complicated: In 
addition to the effects due to higher order operators contributing to the E1 
transition (the dots in Eq.~\eqref{LagE1}), one needs to calculate order $v$ 
and $v^2$ corrections to the initial and final state wave functions due to 
higher order potentials.\footnote{Higher order Fock states become relevant only 
if $\Lambda_{\textrm{QCD}}$ is of the same order as $mv^2$ or larger.} This 
complication has hindered so far complete numerical computations of the E1 
transitions between low-lying heavy quarkonium states within pNRQCD (for 
partial calculations see Refs.~\cite{Pietrulewicz:2013ct,Martinez:2016spe}). 
The present paper aims to close this gap.

The paper is structured in the following way. In Sec.~\ref{sec:theory1} we 
discuss the theoretical background of the computation and display the formulas 
that we use for the decays. In this section, we present also results for the 
electric dipole transitions when only the LO static potential is incorporated 
in the Schr\"odinger equation. Sec.~\ref{sec:theory2} is devoted to present the 
same results but incorporating the complete static potential in the LO 
Hamiltonian. Renormalon effects and resummation of large logarithms are also 
taken into account in this part. All of this leads to a good convergence 
pattern for the studied decay rates and thus to firm predictions for all of 
them. We summarize our results and conclude in Sec.~\ref{sec:conclusion}.

\section{Numerical analysis in pNRQCD at weak coupling: Fixed order calculation}
\label{sec:theory1}

\subsection{Decay width}
\label{subsec:DecayWidth}
We aim at computing electric dipole (E1) transitions from $1P$ bottomonium 
states at order $k_{\gamma}^{3}/m^{2}$ under the condition $mv^2\gg 
\Lambda_{\textrm{QCD}}$. The formulas for the decay widths have been derived in 
Ref.~\cite{Brambilla:2012be}. They read
\begin{widetext}
\begin{align}
\label{eq:FullDecayWidth}
&
\Gamma(n\,^3\!P_J \to n'\,^3\!S_1 + \gamma) = \Gamma_{nn'}^{(0)}\, \Bigg\{ 1 + 
R^{S=1}_{nn'}(J) - \frac{k_\gamma}{6m} - \frac{k_\gamma^2}{60} 
\frac{I_5^{(0)}(n1 \to n'0)}{I_3^{(0)}(n1 \to n'0)} \nonumber \\
& 
\hspace*{2.50cm} + \left[ \frac{J(J+1)}{2} - 2 \right] \Bigg[ 
-\left(1+\kappa_Q^\text{em}\right) \frac{k_\gamma}{2m} + \frac{1}{m^2} (1 + 
2\kappa_Q^\text{em}) \frac{I_2^{(1)}(n1 \to n'0) + 2I_1^{(0)}(n1 \to 
n'0)}{I_3^{(0)}(n1 \to n'0)} \Bigg] \Bigg\} \,, \\
&
\Gamma(n^1\!P_1 \to n'\,^1\!S_0 + \gamma) = \Gamma_{nn'}^{(0)}\, \Bigg\{ 1 + 
R^{S=0}_{nn'} - \frac{k_\gamma}{6m} - \frac{k_\gamma^2}{60} \frac{I_5^{(0)}(n1 
\to n'0)}{I_3^{(0)}(n1 \to n'0)} \Bigg\} \,,
\label{eq:FullDecayWidth2}
\end{align}
\end{widetext}
where $R^{S=1}_{nn'}(J)$ and $R^{S=0}_{nn'}$ include the initial and final 
state corrections due to higher order potentials (see 
Sec.~\ref{subsec:RelativisticCorrectionspot}) and possibly higher order Fock 
states (see Sec.~\ref{subsec:RelativisticCorrectionsFock}). The remaining 
corrections within the curly brackets are the result of taking into account 
additional electromagnetic interaction terms in the Lagrangian suppressed by 
$\O(v^2)$ (the dots in Eq.~\eqref{LagE1}). For completeness, we have displayed 
in the formulas terms proportional to the anomalous magnetic moment, 
$\kappa_{Q}^{\text{em}}$. These terms will, however, not be considered in the 
numerical analyses because they are at least of order  
$\alpha_{\textrm{s}}k_{\gamma}^{3}/m^{2}$ and thus beyond our accuracy.

\begin{figure}[ht]
\includegraphics[width=6truecm]{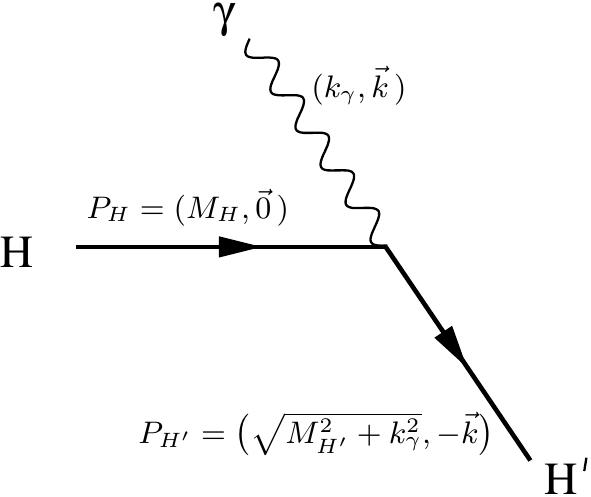}
\caption{\label{fig:kinematics} Kinematics of the radiative transition $H \to 
H^\prime \gamma$ in the rest frame of the initial-state quarkonium $H$.}
\end{figure}

The LO decay width, which scales like $k_{\gamma}^{3}/(mv)^2$, is
\begin{equation}
\Gamma_{nn'}^{(0)} = \frac{4}{9}\, \alpha_\text{em}\, e_Q^2\, k_\gamma^3 
\left[I_3^{(0)}(n1 \to n'0) \right]^{2} \,,
\label{eq:Gamma0}
\end{equation}
with $\alpha_\text{em}$ the electromagnetic fine structure constant, $e_Q$ the 
charge of the heavy quark $Q$ in units of the electron charge, and $k_\gamma$ 
the photon energy determined by the kinematics shown in 
Fig.~\ref{fig:kinematics}:
\begin{equation}
k_{\gamma} = |\vec{k}| = \frac{M_{H}^{2}-M_{H'}^{2}}{2M_{H}} = (M_{H}-M_{H}') + 
\O\left(\frac{k_{\gamma}^{2}}{M_{H}}\right) \,.
\label{kgamma}
\end{equation}
The LO decay width follows from the LO electric dipole interaction in the 
pNRQCD Lagrangian shown in Eq.~\eqref{LagE1}.

All other terms in Eqs.\eqref{eq:FullDecayWidth} and \eqref{eq:FullDecayWidth2} 
are of relative order $v^2$ with respect to the LO decay width. In particular, 
the function
\begin{equation}
I_N^{(k)}(n\l \to n'\l') = \int\limits_0^\infty \d r \, r^{N} 
R_{n'\l'}^{\ast}(r) \left[ \frac{\d^k}{\d r^k} R_{n\l}(r) \right]
\end{equation}
is a matrix element that involves the radial wave functions of the initial and 
final states. From $r \sim 1/p \sim 1/(mv)$ it follows that it scales like 
$(mv)^{2+k-N}$.

Under the assumption $mv^2 \gg \Lambda_{\textrm{QCD}}$ we can compute the 
quarkonium potential in perturbation theory, \textit{i.e.}, as an expansion in 
$\alpha_{\textrm{s}}$. The wave functions are then the solutions of the 
Schr\"odinger equation
\begin{equation}
\label{eq:SchEqu} 
H^{(0)} \psi_{n\l m}(\vec{r}\,) = E_n \psi_{n\l m}(\vec{r}\,) \,,
\end{equation}
where $H^{(0)}$ contains the (perturbative) quark-antiquark static potential.
More specifically, in this section we take the leading order Hamiltonian as 
\begin{equation}
\label{eq:Hamiltonian}
H^{(0)} = -\frac{\vec{\nabla}_r^2}{m} + V_s^{(0)}(r) \,,
\end{equation}
where $-{\vec{\nabla}_r^2}/{m}$ is the (non-relativistic) kinetic energy in the 
center of mass frame and 
\begin{equation}
V_s^{(0)}(r) = - \frac{4\alpha_{\textrm{s}}}{3r} \,.
\label{LOCoulomb}
\end{equation}
This means that we include in the static potential only the LO potential in 
$\alpha_{\textrm{s}}$, which is the Coulomb potential times the Casimir of the 
fundamental representation in SU(3), \textit{i.e.}, $4/3$. A different choice 
will be analyzed in Sec.~\ref{sec:theory2}. With the choice \eqref{LOCoulomb}, 
$\psi_{n\ell m}(\vec{r}\,)$ and $E_{n}$ can be taken from the hydrogen-atom and 
read
\begin{align}
\psi_{n\l m}(\vec{r}\,) &= R_{n\ell}(r) Y_{\l m}(\Omega_r) \nonumber \\
&
= N_{n\l} \, \rho_n^\l \, \e{-\frac{\rho_n}{2}} \, L_{n-\l-1}^{2\l+1}(\rho_n) 
Y_{\l m}(\Omega_r) \,, \label{eq:eigenstate} \\
E_n &= -\frac{4 m \alpha_{\textrm{s}}^2}{9n^2} \,, \label{eq:eigenenergy}
\end{align}
where $\rho_n = 2 r/(n a)$ is a dimensionless variable and $a = 3/(2 m 
\alpha_{\textrm{s}})$ is the Bohr-like radius. The functions 
$L_{n-\l-1}^{2\l+1}$ and $Y_{\l m}$ are the associated Laguerre polynomials and 
the spherical harmonics, respectively. The normalization reads
\begin{equation}
N_{n\l} = \sqrt{\left( \frac{2}{n a} \right)^3 \frac{(n-\l-1)!}{2n[(n+\l)!]}} 
\,.
\end{equation}
Finally, if not differently specified, here and in the rest of the paper, 
$\alpha_{\textrm{s}}$ is understood evaluated at the renormalization scale 
$\nu$: $\alpha_{\textrm{s}} \equiv \alpha_{\textrm{s}}(\nu)$. Hence the 
potential, the Bohr-like radius and, through it, the wave functions depend on 
$\nu$.

\subsection{Relativistic wave function corrections}
\label{subsec:RelativisticCorrections}
The LO wave function \eqref{eq:eigenstate} gets corrections due to higher order 
potentials and possibly to higher order Fock states. Corrections due to higher 
order potentials contribute at relative order $v^2$, and therefore have to be 
included in the analysis to reach a precision of order $k_{\gamma}^{3}/m^{2}$.
These corrections will be outlined in the next 
Sec.~\ref{subsec:RelativisticCorrectionspot}. Corrections due to higher order 
Fock states will be discussed in Sec.~\ref{subsec:RelativisticCorrectionsFock}.

\subsubsection{Corrections due to higher order potentials}
\label{subsec:RelativisticCorrectionspot}
To account for $\mathcal{O}(v^{2})$ corrections to the decay width due to 
higher order potentials, we need to consider the Hamiltonian
\begin{equation}
H = -\frac{\vec{\nabla}_r^2}{m} + V_s(r) + \delta H \,.
\end{equation}
The quark-antiquark static potential up to next-to-next-to-leading order (NNLO) 
is given by
\begin{equation}
V_s(r) = V_s^{(0)}(r) \left[ 1 + \sum\limits_{k=1}^{2} 
\left(\frac{\alpha_{\textrm{s}}}{4\pi}\right)^k a_k(\nu,r) \right] \,,
\label{eq:StatPot}
\end{equation}
where the coefficients of the $\mathcal{O}(\alpha_{\textrm{s}})$ and 
$\mathcal{O}(\alpha_{\textrm{s}}^{2})$ radiative corrections to the LO static 
potential are:
\begin{align}
\label{eq:radiative1}
a_1(\nu,r) &= a_1 + 2\beta_0 \ln{\nu \e{\gamma_E} r} \,, \\
\label{eq:radiative2}
a_2(\nu,r) &= a_2 + \frac{\pi^2}{3} \beta_0^2 + (4a_1 \beta_0 + 2\beta_1) 
\ln{\nu \e{\gamma_E} r} \nonumber \\
& + 4\beta_0^2 \lnSquare{\nu \e{\gamma_E} r} \,, 
\end{align}
with $a_1 = -8+5\beta_0/3$ and $a_2 = 100 n_f^2/81- n_f (1229/27 + 52\zeta_3/3) 
+ 4343/18 + 36 \pi^2 - 9\pi^4/4 + 66 \zeta_3$. The coefficients $\beta_{i}$ are 
the coefficients of the $\beta$-function with $\beta_{0} = 11-2n_f/3$ and 
$\beta_1 = 102 - 38 n_f/3$; $n_{f}$ is the number of massless flavors. The 
$\mathcal{O}(\alpha_{\textrm{s}})$ correction was computed in 
Ref.~\cite{Fischler:1977yf} and the $\mathcal{O}(\alpha_{\textrm{s}}^2)$ one in 
Ref.~\cite{Schroder:1998vy}. In this section, we consider higher order 
corrections to the static potential as perturbations around the leading order 
solution of Sec.~\ref{subsec:DecayWidth}. Hence, the order 
$\alpha_{\textrm{s}}$ correction contributes to the transition width at 
relative order $v$ in first order quantum mechanical perturbation theory and at 
relative order $v^2$ in second order quantum mechanical perturbation theory, 
whereas the order $\alpha_{\textrm{s}}^2$ correction contributes at relative 
order $v^2$ in first order quantum mechanical perturbation theory. On the other 
hand, the $\mathcal{O}(\alpha_{\textrm{s}}^{3})$ correction, which is also 
known from Refs.~\cite{Brambilla:1999qa,Anzai:2009tm,Smirnov:2009fh}, would 
give a contribution to the E1 decay rate of relative order $v^3$, which is 
beyond our precision. Therefore, we will not include 
$\mathcal{O}(\alpha_{\textrm{s}}^{3})$ corrections in this part of our 
analysis.

The term $\delta H$ contains relativistic corrections to the potential and to 
the kinetic energy. They can be organized as an expansion in the inverse of the 
heavy quark mass, $m$. At the order we are interested in, such an expansion 
includes all the $1/m$ and $1/m^2$ potentials and, at order $1/m^{3}$, the 
first relativistic correction to the kinetic energy:
\begin{equation}
\label{eq:deltaH} 
\delta H = -\frac{\vec{\nabla}^2}{4 m} -\frac{\vec{\nabla}_r^4}{4 m^3} + 
\frac{V^{(1)}}{m} + \frac{V_\text{SI}^{(2)}}{m^2} + 
\frac{V_\text{SD}^{(2)}}{m^2} \,.
\end{equation}
At order $1/m^{2}$, we have distinguished between spin-independent (SI) and 
spin-dependent (SD) terms:
\begin{align}
V_\text{SI}^{(2)}(r) &= V_r^{(2)}(r) + \frac{1}{2} \lbrace 
V_{p^2}^{(2)}(r),-\vec{\nabla}_r^2 \rbrace + V_{L^2}^{(2)}(r)\, \vec{L}^2 \,, \\
V_\text{SD}^{(2)}(r) &= V_{LS}^{(2)}(r)\, \vec{L} \cdot \vec{S} + 
V_{S^2}^{(2)}(r)\, \vec{S}^{\;\!2} + V_{S_{12}}^{(2)}(r)\, S_{12} \,,
\end{align}
where $\vec{S} = \vec{S}_{1} + \vec{S}_{2} = (\vec{\sigma}_{1} + 
\vec{\sigma}_{2})/2$, $\vec{L} = \vec{r} \times (-i\vec{\nabla}_r)$ and 
$S_{12} = 3(\hat{r}\cdot\vec{\sigma}_{1}) (\hat{r}\cdot\vec{\sigma}_{2}) - 
\vec{\sigma}_{1}\cdot\vec{\sigma}_{2}$; $\{\;,\;\}$ stands for the 
anticommutator. The above potentials read at leading (non-vanishing) order in 
perturbation theory (see, \textit{e.g.}, Ref.~\cite{Brambilla:2004jw}):
\begin{align}
&
V^{(1)}(r) = -\frac{2 \alpha_{\textrm{s}}^2}{r^2} \,, 
\hspace*{0.40cm}
V_r^{(2)}(r) = \frac{4\pi}{3}  \alpha_{\textrm{s}} \delta^{(3)}(\vec{r}\,) \,, 
\nonumber \\
&
V_{p^2}^{(2)}(r) = -\frac{4\alpha_{\textrm{s}}}{3r} \,,
\hspace*{0.90cm}
V_{L^2}^{(2)}(r) = \frac{2\alpha_{\textrm{s}}}{3 r^3} \,, \nonumber \\
&
V_{LS}^{(2)}(r) = \frac{2\alpha_{\textrm{s}}}{r^3} \,,
\hspace*{1.00cm}
V_{S^2}^{(2)}(r) = \frac{16\pi \alpha_{\textrm{s}}}{9} \delta^{(3)}(\vec{r}\,) 
\,, \nonumber \\
&
V_{S_{12}}^{(2)}(r) = \frac{\alpha_{\textrm{s}}}{3 r^3} \,. 
\end{align}
All these potentials contribute through first order quantum mechanical 
perturbation theory at relative order $v^2$ to the E1 width.

Using quantum-mechanical perturbation theory, we compute the first and, for one 
term, the second order correction, induced by $\delta V=(V_{s}-V_{s}^{(0)}) + 
\delta H$, to the wave function $\psi_{n\l m}(\vec{r}) \equiv \langle \vec{r}| 
n\l m\rangle$ of energy $E_{n}$. The second order correction to the wave 
function is only needed when the perturbation is given by the next-to-leading 
order (NLO) term in the static potential, \textit{i.e.}, the one proportional 
to $a_{1}(\nu,r)$. The (normalized) corrections to the wave function are at 
first order
\begin{equation}
| n\l m\rangle^{(1)} = \sum\limits_{n' \neq n,\l',m'} \frac{\langle n'\l' m'| 
\delta V | n\l m\rangle}{E_n - E_{n'}} | n'\l' m'\rangle\,,
\label{eq:1st}
\end{equation}
and at second order 
\begin{widetext}
\begin{align}
\label{eq:2nd}
| n\l m\rangle^{(2)} &= \!\!\!\!\! \sum\limits_{k_{1} \neq n,\l_1,m_1} \Bigg[ 
\sum\limits_{k_{2} \neq n,\l_2,m_2} \!\!\!\!\! \frac{\langle k_{1}\l_1 
m_1|\delta  V | k_{2}\l_2 m_2\rangle \langle k_{2}\l_2 m_2|\delta  V | n\l 
m\rangle}{(E_{n} - E_{k_{1}}) (E_{n} - E_{k_{2}})} - \frac{\langle k_{1}\l_1 
m_1|\delta  V | n\l m\rangle \langle n\l m|\delta  V | n\l m\rangle}{(E_{n} - 
E_{k_{1}})^{2}} \Bigg] | k_{1}\l_1 m_1\rangle \nonumber \\
&
- \frac{1}{2} \! \sum\limits_{k_{2} \neq n,\l_2,m_2} \!\!\!\! \frac{|\langle 
k_{2}\l_2 m_2| \delta V | n\l m\rangle|^2}{(E_{n} - E_{k_{2}})^{2}} | n\l 
m\rangle \,.
\end{align}
The operator $\displaystyle \sum\limits_{n' \neq n,\l',m'} \frac{| n'\l' 
m'\rangle \langle n'\l' m'|}{E_{n} - E_{n'}}$ appearing in the 
Eqs.~\eqref{eq:1st} and \eqref{eq:2nd} can be rewritten as
\begin{equation}
\lim\limits_{ E \to E_n} \left( \sum\limits_{n',\l',m'} \! \frac{| n'\l' 
m'\rangle \langle n'\l' m'|}{E - E_{n'}} - \sum\limits_{\l',m'} \! \frac{| n\l' 
m'\rangle \langle n\l' m'|}{E - E_{n}} \right) \equiv \frac{1}{(E_n - H)'} \,,
\end{equation}
and it can thus be identified with the pole-subtracted Coulomb Green function. 
In coordinate space, it reads
\begin{equation}
\label{eq:GreenFunctionI}
G'_n(\vec{r}_1,\vec{r}_2) \equiv  \langle\vec{r}_1|\frac{1}{(E_n - 
H)'}|\vec{r}_2\rangle = \lim_{ ~ E \to E_n} \left(G(\vec{r}_1,\vec{r}_2) - 
\sum\limits_{\l=0}^{n-1} \sum\limits_{m=-\l}^{\l} \frac{\psi^*_{n\l 
m}(\vec{r}_1) \psi_{n\l m}(\vec{r}_2)}{E - E_n}\right),
\end{equation}
where $G(\vec{r}_1,\vec{r}_2)$ is the Coulomb Green 
function~\cite{Voloshin:1979uv,Kiyo:2014uca}:
\begin{equation}
\label{eq:GreenFunctionII}
G(\vec{r}_1,\vec{r}_2) = - \sum\limits_{\l=0}^\infty \frac{2\l+1}{4\pi} 
P_\l(\hat{r}_1 \cdot \hat{r}_2) G_\l(r_1,r_2), \quad \textrm{with} \quad 
G_\l(r_1,r_2) = \sum\limits_{\nu=\l+1}^\infty \frac{m}{2} a^2 
\left(\frac{\nu^4}{\lambda}\right) 
\frac{R_{\nu\l}(\r{\lambda}{1})R_{\nu\l}(\r{\lambda}{2})}{\nu-\lambda},
\end{equation}
$E \equiv -{4m \alpha_{\textrm{s}}^{2}}/{(9\lambda^{2})}$ and $\r{\lambda}{i} = 
2r_i/(\lambda a)$. In calculations it may be useful to set $\lambda = n/\sqrt{1 
- \epsilon}$, since in this way we have $E = E_{n} (1 - \epsilon)$ and $E \to 
E_n$ for $\epsilon \to 0$. Therefore, the first order and second order 
corrections to the expectation values of an arbitrary operator $O$ may be 
written as (note that, for the sake of simplicity, only initial state 
corrections are shown herein, but the same corrections affect also the final 
state):
\begin{align}
\langle n'\l' m'| O | n\l m\rangle^{(1)} 
&= \int \d^3 r_1 \, \d^3 r_2 \, \psi_{n'\l'm'}^*(\vec{r}_2) \, O(\vec{r}_2) 
\, G'_n(\vec{r}_2,\vec{r}_1) \, \delta V(\vec{r}_1) \, \psi_{n\l m}(\vec{r}_1) 
\,,
\label{eq:1st2nd}
\\
\langle n'\l' m'| O | n\l m\rangle^{(2)} 
&= \int \d^3 r_1 \, \d^3 r_2 \, \d^3 r_3 \, \psi_{n'\l'm'}^*(\vec{r}_3) \, 
O(\vec{r}_3) \, G'_n(\vec{r}_3,\vec{r}_2) \, \delta V(\vec{r}_2) \, 
G'_n(\vec{r}_2,\vec{r}_1) \,\delta  V(\vec{r}_1) \psi_{n\l m}(\vec{r}_1) 
\nonumber \\
& \hspace{-2.5cm}
- \delta E_V^{(1)}\, \int \d^3 r_1 \, \d^3 r_2 \, \d^3r_3 \, 
\psi_{n'\l'm'}^*(\vec{r}_3) \, O(\vec{r}_3) \, G'_n(\vec{r}_3,\vec{r}_2) 
\, G'_n(\vec{r}_2,\vec{r}_1) \, \delta V(\vec{r}_1) \psi_{n\l m}(\vec{r}_1) 
\nonumber \\
& \hspace{-2.5cm}
- \frac{1}{2} \int \d^3 r \, \psi_{n'\l' m'}^*(\vec{r}) \, O(\vec{r}) \, 
\psi_{n\l m}(\vec{r}) \int \d^3 r_1 \, \d^3 r_2 \, \d^3 r_3 \, 
\psi_{n\l m}^*(\vec{r}_3) \, \delta V(\vec{r}_3) \, G'_n(\vec{r}_3,\vec{r}_2) 
\, G'_n(\vec{r}_2,\vec{r}_1) \, \delta V(\vec{r}_1) \psi_{n\l m}(\vec{r}_1) ,
\label{eq:2nd2nd}
\end{align}
where $\delta E_V^{(1)}$ is the first order correction to the energy: 
$\displaystyle \delta E_V^{(1)} \equiv \int \d^3 r \, \psi_{n\l m}^*(\vec{r}\,) 
\, \delta V(\vec{r}\,) \, \psi_{n\l m}(\vec{r}\,)$.
\end{widetext}
As a final remark we note that, although in \eqref{eq:deltaH} we have included 
the center of mass kinetic energy, $-{\vec{\nabla}^2}/{4 m}$, this term does 
not contribute at our accuracy.\footnote{Potentials depending on the center of 
mass momentum contribute, instead, at relative order $v^2$ to M1 
transitions~\cite{Brambilla:2005zw}.} The reason is that, even if the center of 
mass kinetic energy scales like a term of relative order $v^2$, nevertheless, 
its contribution vanishes at first order in quantum mechanical perturbation 
theory, Eq.~\eqref{eq:1st}, as the states are eigenstates (in fact simple plane 
waves) of the center of mass momentum.

\subsubsection{Corrections due to higher order Fock states}
\label{subsec:RelativisticCorrectionsFock}
The LO correction to E1 transitions due to higher order Fock states comes 
from diagrams in which a heavy quark-antiquark color singlet state is coupled 
to a heavy quark-antiquark color octet state via emission and reabsorption of 
gluons whose energy and momentum are of order $m v^2$ or 
$\Lambda_{\textrm{QCD}}$. The coupling of the color singlet field S with the 
color octet field O is encoded in the pNRQCD Lagrangian in a chromoelectric 
dipole interaction term: $\displaystyle \int d^3 r \; \mathrm{Tr} \left\{ 
\mathrm{S}^{\dagger} {\vec{r}}\cdot g {\vec{E}} \mathrm{O}  + \mathrm{H.c} 
\right\}$. The relevant Feynman diagrams in pNRQCD are shown in Fig.~8 of 
Ref.~\cite{Brambilla:2012be}: they are diagrams corresponding to the 
normalization of the initial and final state wave functions, diagrams 
accounting for the corrections to the initial and final state wave functions 
due to the presence of octet states, and a diagram representing an electric 
dipole transition mediated by an intermediate octet state. According to the 
power counting of pNRQCD, those diagrams contribute to relative order
$\alpha_{\textrm{s}}v^{2}$ if the gluons carry an energy and a momentum of 
order $m v^2$. They contribute to relative order 
$\Lambda_{\text{QCD}}^{2}/(mv)^{2}$ or $\Lambda_{\text{QCD}}^{3}/(m^{3}v^{4})$ 
if the gluons are non-perturbative and carry an energy and a momentum of order 
$\Lambda_{\textrm{QCD}}$. In the first case, their contribution is smaller than 
$v^2$ by a factor $\alpha_{\textrm{s}}$ and hence beyond our accuracy. In the 
second case, it is also smaller than $v^2$ if $mv^{2} \gg 
\Lambda_{\text{QCD}}$, which is what we have assumed. It should be remarked, 
however, that it suffices $mv^2 \sim \Lambda_{\text{QCD}}$ for the 
non-perturbative contributions to be of the same relative order, $v^2$, as the 
ones coming from higher order potentials.

\subsection{Numerical analysis}
We specify, first, the parameters that enter in the determination of the 
bottomonium E1 transition widths. We have
\begin{equation}
n_f = 3 , \hspace*{0.2cm} e_{b} = -\frac{1}{3}, \hspace*{0.2cm} 
\alpha_{\text{em}} = \frac{e^2}{4\pi} \approx \frac{1}{137} \,,
\end{equation}
where $n_f$ is the number of massless flavors,\footnote{At the typical momentum 
transfer inside the $b\bar{b}$ system the charm quark 
decouples~\cite{Brambilla:2001qk}.} $e_{b}$ is the electric charge of the 
bottom quark in units of the electron charge $e$ and $\alpha_\text{em}$ is the 
electromagnetic fine structure constant.

\begin{table}[!t]
\centering
\begin{tabular}{rcccccc}
\hline
\hline
Notation & $\eta_b(1S)$ & $\Upsilon(1S)$ & $h_b(1P)$ & $\chi_{b0}(1P)$ & 
$\chi_{b1}(1P)$ & $\chi_{b2}(1P)$ \\
$n\,^{2s+1}\!\l_J$ & $1\,^1\!S_0$ & $1\,^3\!S_1$ & $2\,^1\!P_1$ & $2\,^3\!P_0$ 
& $2\,^3\!P_1$ & $2\,^3\!P_2$ \\
\hline
Mass~\cite{PhysRevD.98.030001} & $9.399$ & $9.460$ & $9.899$ & $9.859$ & 
$9.893$ & $9.912$ \\
\hline
\hline
\end{tabular} 
\caption{\label{tab:QuarkoniaMasses} Masses in GeV of the bottomonium states 
involved in the electric dipole transitions considered in this work, from the 
PDG~\cite{PhysRevD.98.030001}.}
\end{table}

\begin{figure*}[!t]
\centering
\includegraphics[width=0.98\textwidth]{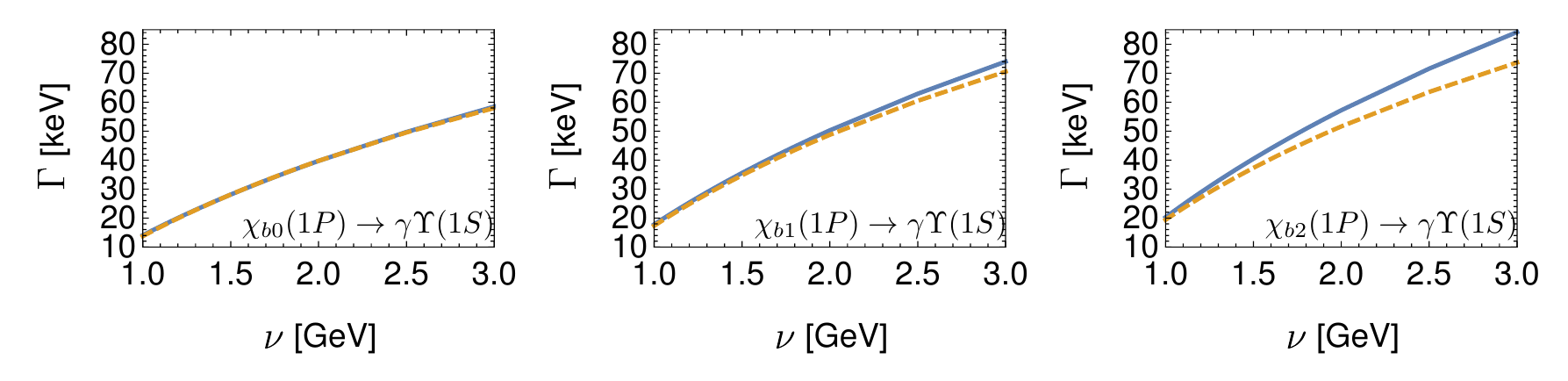}
\caption{\label{fig:23PJ13S1-relativistic}
For the electric dipole transitions $\chi_{bJ}(1P)\to \gamma\Upsilon(1S)$, with 
$J=0$ (left panel), $J=1$ (middle panel) and $J=2$ (right panel), we show the 
leading order decay rate (solid blue curve) and the decay rate obtained 
including the contributions in Eq.~\eqref{eq:FullDecayWidth} that stem from 
higher order electromagnetic operators (dashed orange curve).}
\end{figure*}

\begin{figure*}[!t]
\centering
\includegraphics[width=0.98\textwidth]{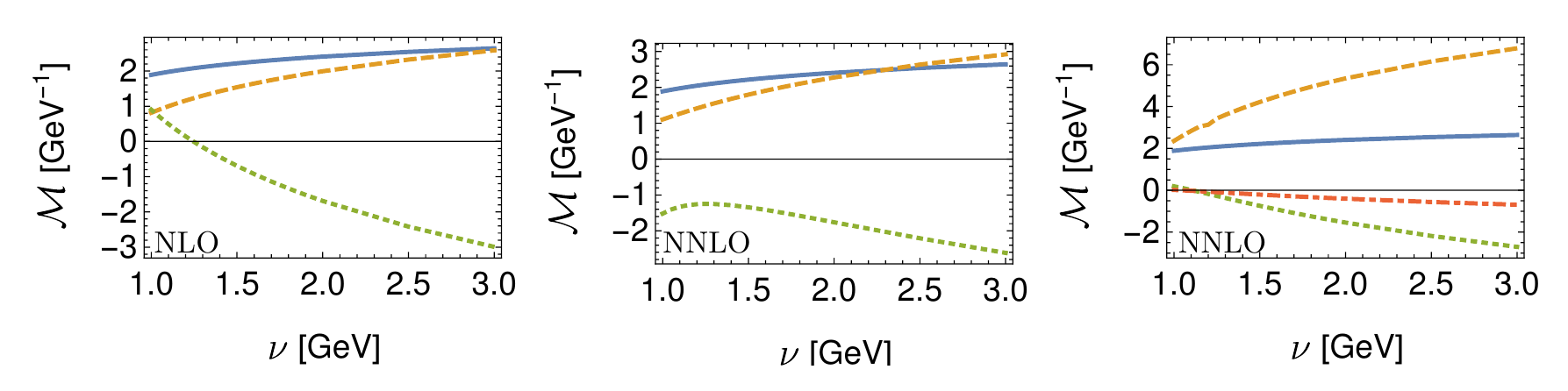}
\caption{\label{fig:23PJ13S1-static}  Matrix elements of the one and two loop 
corrections to the static potential contributing to the decay width 
$\chi_{bJ}(1P)\to \gamma\Upsilon(1S)$. In all panels, the solid blue line 
indicates the LO (no loop corrections) matrix element, the dashed orange line 
indicates the initial state correction and the dotted green line indicates the 
final state correction. \textit{Left panel:} First order correction to the 
decay width due to the NLO static potential (the term proportional to 
$a_1(\nu,r)$). \textit{Middle panel:} First order correction to the decay width 
due to the NNLO static potential (the term proportional to $a_2(\nu,r)$). 
\textit{Right panel:} Second order correction to the decay width due to the NLO 
static potential. The additional dot-dashed red line corresponds to a matrix 
element with a first order correction to both the initial and final states. The 
matrix elements do not depend on $J$.}
\end{figure*}

The masses of the initial and final quarkonium states are chosen to be the ones 
reported by the PDG~\cite{PhysRevD.98.030001}, listed in 
Table~\ref{tab:QuarkoniaMasses}. The photon energies are determined by the 
kinematics of the two body decay, Eq.~\eqref{kgamma}, and are given by
\begin{equation}
\begin{split}
k_\gamma &
= \left\lbrace \begin{array}{ll}
391.1~\text{MeV} & \mbox{ for } \chi_{b0}(1P) \to \Upsilon(1S) + \gamma \,, \\
423.0~\text{MeV} & \mbox{ for } \chi_{b1}(1P) \to \Upsilon(1S) + \gamma \,, \\
441.6~\text{MeV} & \mbox{ for } \chi_{b2}(1P) \to \Upsilon(1S) + \gamma \,, \\ 
488.3~\text{MeV} & \mbox{ for }\hspace*{0.15cm} h_b(1P) \to \eta_b(1S) + \gamma 
\,.
\end{array} \right.
\end{split}
\label{kgammavalues}
\end{equation}
Our reference value for the strong coupling constant is 
$\alpha_{\textrm{s}}^{(n_{f}=3)}(1\,\text{GeV})=0.480$. We obtain this value by 
using the \texttt{RunDec} package~\cite{Chetyrkin:2000yt} to run down from 
$\alpha_{\textrm{s}}^{(n_{f}=5)}(M_{Z} = 91.19\,\textrm{GeV})=0.118$ at 
four-loop accuracy. We then run $\alpha_{\textrm{s}}$ to the typical scales of 
the bound state.

We fix the bottom quark pole mass using the experimental mass of the 
$\Upsilon(1S)$ state and the leading order binding energy. This means that if
\begin{equation}
M_\text{exp.}(\Upsilon(1S)) = 2 m - \frac{4m\alpha_{\textrm{s}}^2}{9} + 
\O(\alpha_{\textrm{s}}^3) \,,
\end{equation}
the bottom mass is
\begin{equation}
m = \frac{M_\text{exp.}(\Upsilon(1S))}{2} \left(1 + 
\frac{2\alpha_{\textrm{s}}^2}{9} + \O(\alpha_{\textrm{s}}^3)\right) \,,
\end{equation}
which is the expression that goes into the wave function. Higher order terms 
are beyond our accuracy. Indeed, even the $\O(\alpha_{\textrm{s}}^2)$ term 
given above is beyond our accuracy if used for higher order corrections in the 
$1/m$ expansion. For those corrections we set the bottom quark mass to be
\begin{equation}
m = \frac{M_\text{exp.}(\Upsilon(1S))}{2} \,.
\end{equation}

\begin{figure*}[!t]
\centering
\includegraphics[width=0.98\textwidth]{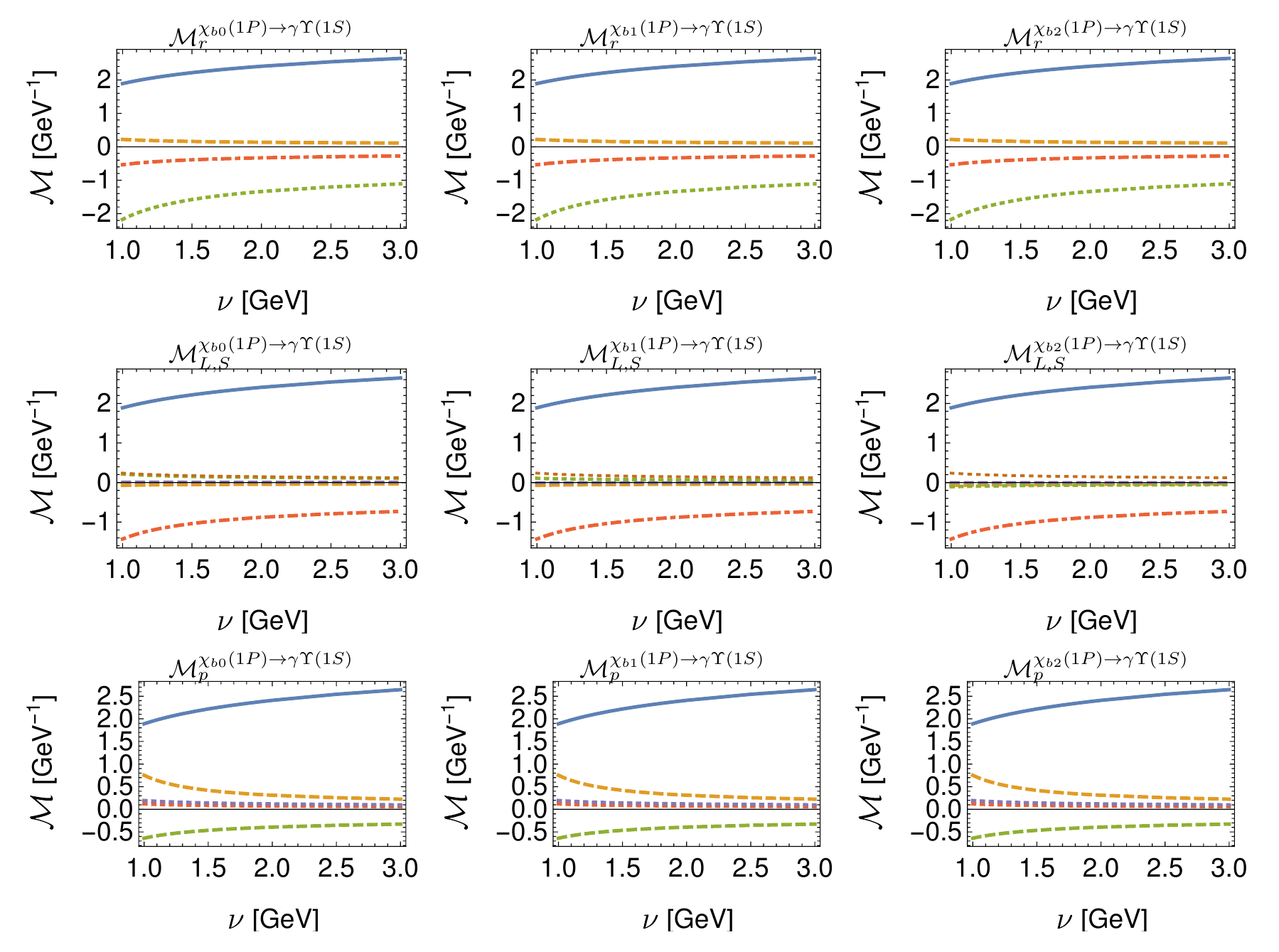}
\caption{\label{fig:23PJ13S1-others}  For the electric dipole transitions 
$\chi_{bJ}(1P)\to \gamma\Upsilon(1S)$, with $J=0$ (first column), $J=1$ (second 
column) and $J=2$ (third column), we show the matrix elements contributing to 
the decay widths induced by the LO electric dipole operator \eqref{LagE1} when 
the bottomonium wave functions include relativistic corrections due to higher 
order potentials and kinetic energy in the $1/m$ expansion, see 
Sec.~\ref{subsec:RelativisticCorrectionspot}. \textit{First row:} Matrix 
element at LO, \textit{i.e.}, without relativistic corrections, (solid blue), 
and at NNLO: Including relativistic corrections due to the $V^{(1)}$ potential 
(dashed orange and dotted green lines for initial and final state corrections, 
respectively) and due to the $V_r^{(2)}$ potential (dot-dashed red line for 
final state correction). \textit{Second row:} Matrix element at LO (solid blue),
and at NNLO: Including relativistic corrections due to the $V_{L^2}^{(2)}$ 
potential (dashed orange for initial state correction), due to the 
$V_{LS}^{(2)}$ potential (dotted green for initial state correction), due to 
the $V_{S^2}^{(2)}$ potential (dot-dashed red for final state correction), and 
due to the $V_{S_{12}}^{(2)}$ potential (dashed violet and dotted brown for 
initial and final state corrections, respectively). \textit{Third row:} Matrix 
element at LO (solid blue), and at NNLO: Including relativistic corrections due 
to the $V_{p^{2}}^{(2)}$ potential (dashed orange and green for initial and 
final state corrections, respectively), and due to the kinetic energy term 
$-\vec{\nabla}^{4}/4m^{3}$ (dotted red and violet for initial and final state 
corrections, respectively).}
\end{figure*}

\subsubsection{\texorpdfstring{$\chi_{bJ}(1P)\to \gamma\Upsilon(1S)$ with $J=0,1,2$}{chibJ(1P)to gamma Upsilon(1S) with J=0,1,2}}
We begin the numerical analysis of the electric dipole transitions 
$\chi_{bJ}(1P)\to \gamma\Upsilon(1S)$, with $J=0,1,2$, focusing on the 
contributions that appear in Eq.~\eqref{eq:FullDecayWidth} and come from higher 
order electromagnetic operators in the pNRQCD Lagrangian. As one can see in 
Fig.~\ref{fig:23PJ13S1-relativistic}, the leading order decay width depends 
strongly on the renormalization scale $\nu$. This is due to the scale 
dependence of the Bohr-like radius that enters the wave functions. The effects 
from higher order electromagnetic operators are small. The correction to the LO 
decay width is at most $\approx 1\%$, $\approx 2\%$ and $\approx 5\%$ when the 
initial state is a $\chi_{b0}$, $\chi_{b1}$ and $\chi_{b2}$, respectively. This 
can be understood analyzing each contribution separately: The contributions 
almost cancel for $J=0$ but this is not the case for $J=1$ and $J=2$.

\begin{figure*}[!t]
\centering
\includegraphics[width=0.98\textwidth]{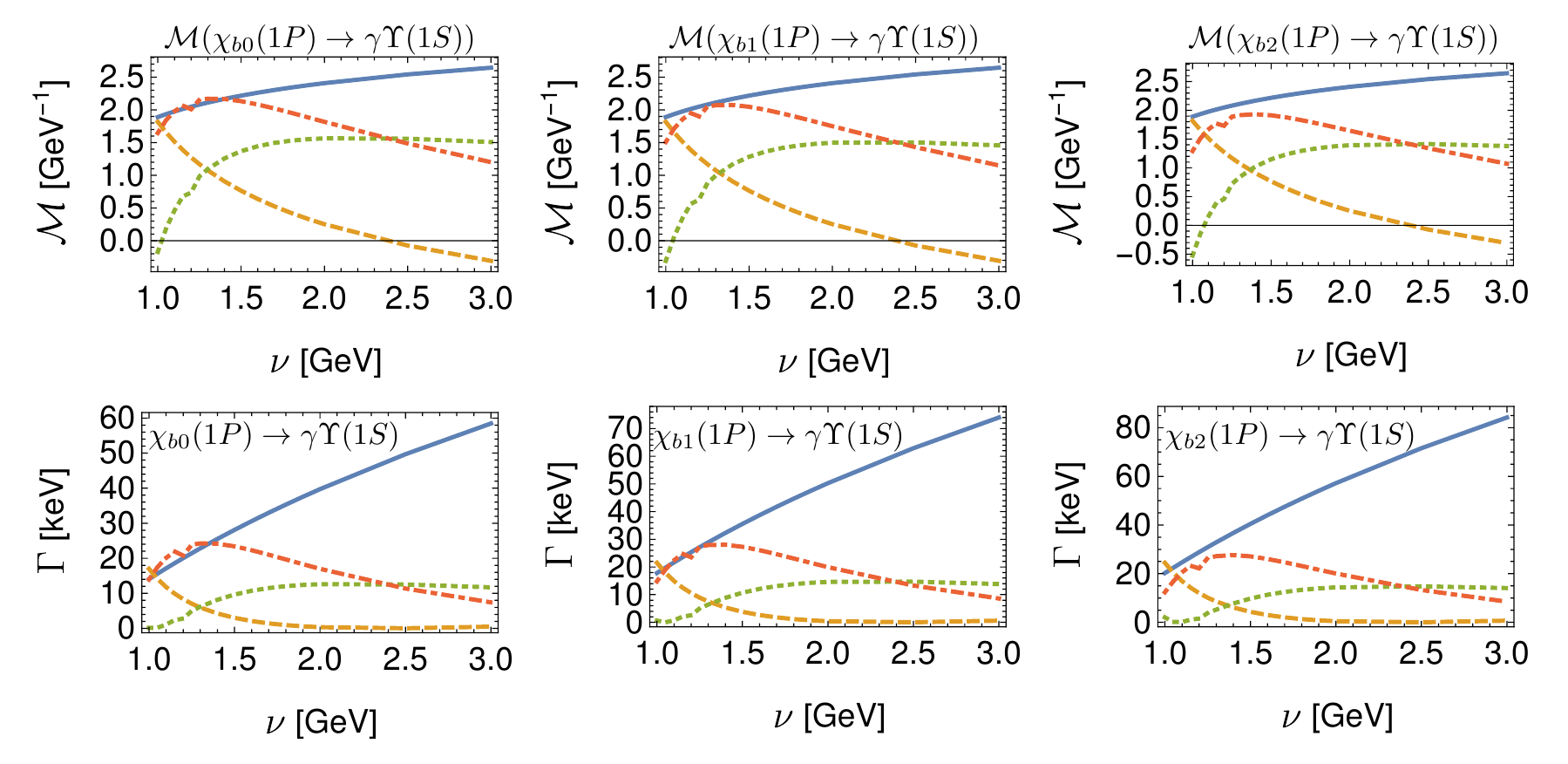}
\caption{\label{fig:23PJ13S1-orderBYorder} Sums of the  matrix elements shown 
in Figs.~\ref{fig:23PJ13S1-static} and~\ref{fig:23PJ13S1-others}, and decay 
widths at different orders for the E1 transitions $\chi_{bJ}(1P)\to 
\gamma\Upsilon(1S)$. The three panels in each row refer to the three cases 
$J=0,1,2$, respectively. \textit{First row:} Sum of the matrix elements at (not 
up to) LO (solid blue), NLO (dashed orange), NNLO (dotted green) and NLO+NNLO 
(dot-dashed red). \textit{Second row:} Decay widths at (not up to) LO (solid 
blue), NLO (dashed orange), NNLO (dotted green) and NLO+NNLO (dot-dashed red).}
\end{figure*}

The radiative corrections to the LO static potential (the terms in the sum of 
Eq.~\eqref{eq:StatPot}) lead to first order and second order quantum-mechanical 
corrections to the decay widths. These terms are proportional to (soft) logs 
(like $\lnEne{\nu r}$) and thus one expects a significant scale dependence of 
the resulting matrix elements. This is indeed the case as shown in 
Fig.~\ref{fig:23PJ13S1-static}.\footnote{Since these are central potentials, 
the matrix elements do not depend on the spin $s$ or the total angular momentum 
$J$ and thus the result is the same for all $P$-wave to $S$-wave transitions.} 
The plotted matrix elements, $\M$, stand for the first order and second order 
corrections to the matrix elements of the specified potentials, according to 
Eqs.~\eqref{eq:1st2nd} and~\eqref{eq:2nd2nd}. The left and middle panels refer 
to the first order initial and final wave function corrections coming from 
$a_1(\nu,r)$ and $a_2(\nu,r)$, respectively. The right panel refers to the 
second order correction due to the $a_1(\nu,r)$ term of the static potential. 
Among the features shown by the panels, the following are of particular 
interest: \textit{(i)} The matrix elements clearly exceed the value of the LO 
one. To some extent, this is due to the factors stemming from the 
$\beta$-function in Eqs.~\eqref{eq:radiative1} and \eqref{eq:radiative2} that 
are large. \textit{(ii)} The matrix elements depend strongly on the scale 
$\nu$, especially for small~$\nu$. A similar behavior shows up in some matrix 
elements contributing to the M1 transitions~\cite{Pineda:2013lta}. 
\textit{(iii)} The zero crossing in some of the matrix elements comes from the 
logarithms in the Eqs.~\eqref{eq:radiative1} and~\eqref{eq:radiative2}. The 
scale where this effect occurs is $\nu\approx 1.2\,\text{GeV}$. \textit{(iv)} 
Initial and final state corrections partially cancel each other, order by order.

The corrections to the matrix element of the LO electric dipole operator 
\eqref{LagE1}, due to the relativistic corrections to the bottomonium wave 
functions discussed in Sec.~\ref{subsec:RelativisticCorrectionspot}, are shown 
in Fig.~\ref{fig:23PJ13S1-others}. These corrections contribute to the term 
$R_{21}^{S=1}(J)$ in Eq.~\eqref{eq:FullDecayWidth}. As one can see, most of the 
contributions are small, except for the final state correction induced by 
$V^{(1)}$ and the correction due to $V_{S^2}$. The overall dependence on the 
scale $\nu$ is weak in all cases but a slight trend towards larger values by 
decreasing scale can be observed.

\begin{figure*}[!t]
\centering
\includegraphics[width=0.98\textwidth]{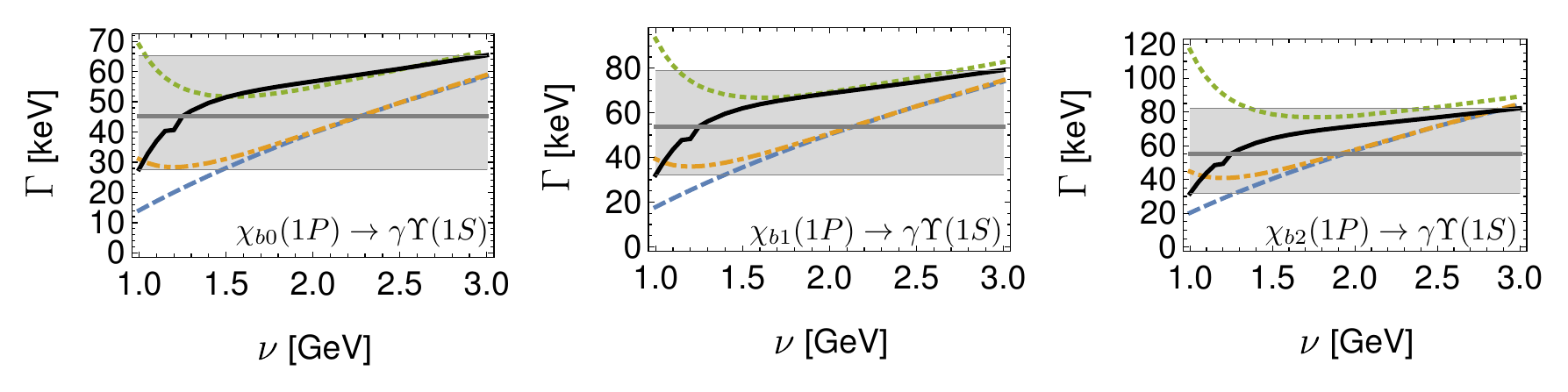}
\caption{\label{fig:23PJ13S1-FinalResult} Decay widths of the electric dipole 
transitions $\chi_{bJ}(1P)\to \gamma\Upsilon(1S)$. The three panels refer to 
the three cases $J=0,1,2$, respectively. In each panel, the dashed blue curve 
is the LO decay width, the dot-dashed orange one incorporates LO + NLO 
corrections and the solid black curve incorporates LO + NLO + NNLO corrections 
coming from higher order electromagnetic operators, radiative corrections to 
the static potential and higher order relativistic corrections to the potential 
and the kinetic energy. The dotted green curve is similar to the black one but 
it omits all corrections to the decay width due to radiative corrections of the 
static potential (one and two loops). We take our central value at $\nu = 
1.25\,\text{GeV}$, whereas the gray band indicates the associated uncertainty. 
The scale setting and the uncertainty estimate are explained in the text.}
\end{figure*}

We sum the matrix elements that include radiative corrections to the static 
potential (see Fig.~\ref{fig:23PJ13S1-static}) and higher order relativistic 
corrections to the potential and kinetic energy (see 
Fig.~\ref{fig:23PJ13S1-others}) at each order and the result is shown in 
Fig.~\ref{fig:23PJ13S1-orderBYorder}, first row. The corresponding decay widths 
are displayed in the second row. From both plot sequences we can see that each 
LO, NLO and NNLO contribution depends strongly on the renormalization scale and 
also that subleading contributions may be of similar size to the leading one.
Moreover, the overall impact of the corrections decreases with increasing total 
angular momentum: For $J=0$ the NLO+NNLO  curves exceed for some $\nu$ the LO 
curve, for $J=1$ they touch each other and for $J=2$ they stay slightly below.
The kink, visible in the NNLO matrix element (and subsequently also in the 
NLO+NNLO matrix element) at about $1.2\,\text{GeV}$, can be traced back either 
to the zero crossing or to the maximum in the matrix elements of 
Fig.~\ref{fig:23PJ13S1-static}. Also the NLO and NNLO matrix element sums show 
a zero crossing, and the combined NLO+NNLO matrix element has a clear maximum. 
The zero crossings yield vanishing contribution to the respective decay widths, 
as visible in the second row.

The results that follow from summing up all previous corrections, \textit{i.e.},
those that contribute to the term $R^{S=1}_{21}(J)$ in \eqref{eq:FullDecayWidth}
(we recall that these are radiative corrections to the static potential, due to 
the one and two loop corrections in \eqref{eq:StatPot}, and higher order 
relativistic corrections to the potential and the kinetic energy, due to 
\eqref{eq:deltaH}; their combined effect to the E1 transition width is shown in 
Fig.~\ref{fig:23PJ13S1-orderBYorder}) and those that contribute to the terms 
other than $R^{S=1}_{21}(J)$ in \eqref{eq:FullDecayWidth} (these are due to 
higher order electromagnetic operators in the pNRQCD Lagrangian \eqref{LagE1};
their effect to the E1 transition width is shown in 
Fig.~\ref{fig:23PJ13S1-relativistic}), are shown in 
Fig.~\ref{fig:23PJ13S1-FinalResult}. The renormalization scale dependence of 
the decay widths is reduced as the NLO and NNLO corrections are included.
For instance, varying the renormalization scale from 1~GeV to 3~GeV for the 
$J=1$ case, the LO spans over the range $(17$-$74)\,\text{keV}$, incorporating 
the NLO contribution shrinks the range to $(35$-$75)\,\text{keV}$, and adding 
the NNLO corrections results in a range of $(32$-$79)\,\text{keV}$. Although a 
slight shift towards higher upper bounds is noticeable, the whole range and 
thus the overall scale dependence somewhat decreases from the LO.

Another feature of the panels in Fig.~\ref{fig:23PJ13S1-FinalResult} is that by 
setting the terms proportional to $a_{1}(\nu,r)$ and $a_{2}(\nu,r)$ to zero, 
the decay width exhibits a different $\nu$-dependence in the low $\nu$ region 
(dotted green curve). This suggests that the terms proportional to the logs in 
Eqs.~\eqref{eq:radiative1} and \eqref{eq:radiative2} give rise to 
non-negligible contributions, whose dependence on the renormalization scale 
needs to be treated carefully, as we shall see in the next section.

The convergence of the perturbative series is poor. This can be seen by looking 
at the difference between the LO and NLO, and between the NLO and NNLO results.
Also the strong scale dependence in the range 1~GeV $\le \nu \le$ 3~GeV is a 
consequence of having large higher order corrections. As a consequence, it is 
difficult (if not impossible) to get a reliable result using fixed order 
perturbation theory. Nevertheless, in the following we will produce a first 
rough determination of the $1P$ bottomonium dipole electric transitions,
with a large error reflecting the large uncertainty. We will overcome this 
difficulty and provide a reliable determination with a small uncertainty in the 
next section.

We choose to set the central value of the decay widths at the renormalization 
scale that self-consistently solves the Bohr-like radius equation:
\begin{equation}
\frac{1}{a} = \frac{2m \alpha_{\textrm{s}}(1/a)}{3} \,.
\label{Bohrradius}
\end{equation}
This scale is $\nu = 1/a = 1.25\,\text{GeV}$.\footnote{This is the typical 
momentum transfer inside $n=1$ bottomonia and the largest, most relevant, scale 
in the E1 matrix elements. In particular, it is larger than the typical 
momentum transfer inside $n=2$ bottomonia. The possibility for a 
renormalization scale as low as 1~GeV is accounted for in the uncertainties.}

We estimate the uncertainty associated to the central value in a twofold way: 
\textit{(i)} First, we vary the renormalization scale from 1~GeV to 3~GeV, which
is a conservative interval including the lowest scale where perturbation theory 
may be still applicable and more than twice the inverse of the Bohr radius.
\textit{(ii)} Second, we estimate the uncertainty associated with truncating 
the perturbative series at NNLO and the fact that the series is poorly 
converging by taking one half of the maximum difference between the LO and the 
NNLO decay width. For the final error we choose the largest of these two 
values, which is indicated in the plots by a gray band. Further sources of 
uncertainties are given by the input parameters, these being the masses of 
initial and final states, and the value of $\alpha_{\textrm{s}}$. If we assume 
that these quantities are accurate within $\lesssim (1$-$3)\%$, their 
uncertainty is largely inside the final error.

Hence, a fixed order determination at NNLO gives for the E1 transition widths 
of the $\chi_{bJ}(1P)$: 
\begin{align}
& \Gamma(\chi_{b0}(1P)\to \gamma\Upsilon(1S)) = 45^{+20}_{-18}~\text{keV} \,, \\
& \Gamma(\chi_{b1}(1P)\to \gamma\Upsilon(1S)) = 54^{+25}_{-22}~\text{keV} \,, \\
& \Gamma(\chi_{b2}(1P)\to \gamma\Upsilon(1S)) = 55^{+27}_{-24}~\text{keV} \,.
\end{align}
As anticipated, the errors are large, reflecting the poor convergence of the 
perturbative series. In Sec.~\ref{sec:theory2}, we will see how resumming the 
known terms of the perturbative expansion of the static potential into the wave 
functions will enormously improve the above determinations providing convergent 
expansions with tiny theoretical uncertainties.

\begin{figure}[!t]
\centering
\includegraphics[width=0.45\textwidth]{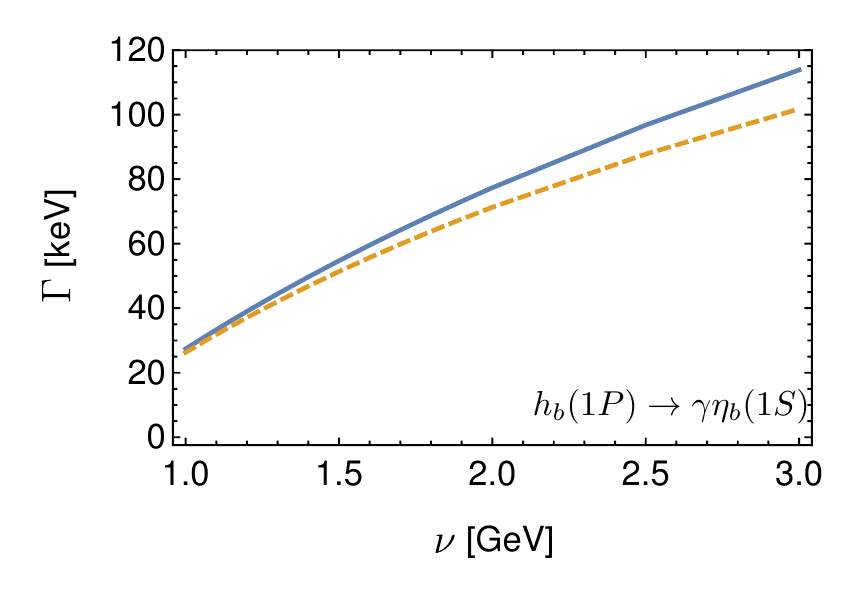}
\caption{\label{fig:21P111S0-relativistic} For the electric dipole transition 
$h_b(1P)\to \gamma\eta_b(1S)$, we show the decay rate at leading order (solid 
blue curve) and including the contributions in Eq.~\eqref{eq:FullDecayWidth2} 
that stem from higher order electromagnetic operators (dashed orange curve).}
\end{figure}

\begin{figure*}[!t]
\centering
\includegraphics[width=0.98\textwidth]{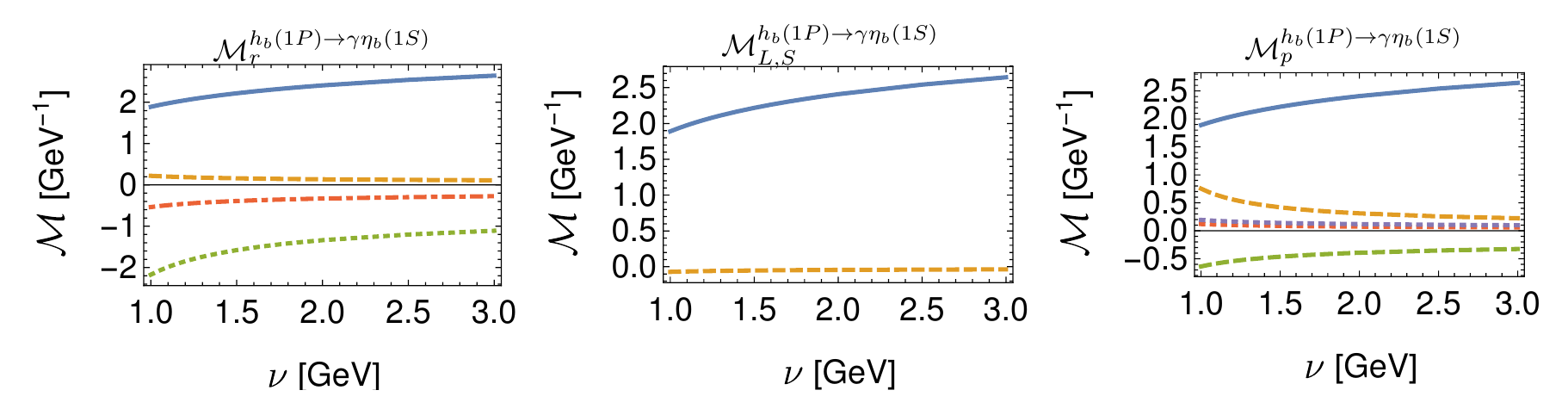}
\caption{\label{fig:21P111S0-others} For the electric dipole transition 
$h_{b}(1P)\to \gamma\eta_{b}(1S)$, we show the matrix elements contributing to 
the decay width induced by the LO electric dipole operator \eqref{LagE1} when 
the bottomonium wave functions include relativistic corrections due to higher 
order potentials and kinetic energy in the $1/m$ expansion, see 
Sec.~\ref{subsec:RelativisticCorrectionspot}. \textit{Left panel:} Matrix 
elements at LO (solid blue), and at NNLO: including relativistic corrections 
due to the $V^{(1)}$ potential (dashed orange and dotted green lines for 
initial and final state corrections, respectively) and due to the $V_r^{(2)}$ 
potential (dot-dashed red line for final state correction). \textit{Middle 
Panel:} Matrix element at LO (solid blue), and at NNLO including relativistic 
corrections due to the $V_{L^2}^{(2)}$ potential (dashed orange for initial 
state correction). \textit{Right panel:} Matrix element at LO (solid blue), and 
at NNLO: including relativistic corrections due to the $V_{p^{2}}^{(2)}$ 
potential (dashed orange and green for initial and final state corrections, 
respectively), and due to the kinetic energy term $-\vec{\nabla}^{4}/4m^{3}$ 
(dotted red and violet for initial and final state corrections, respectively).}
\end{figure*}

\begin{figure*}[!t]
\centering
\includegraphics[width=0.98\textwidth]{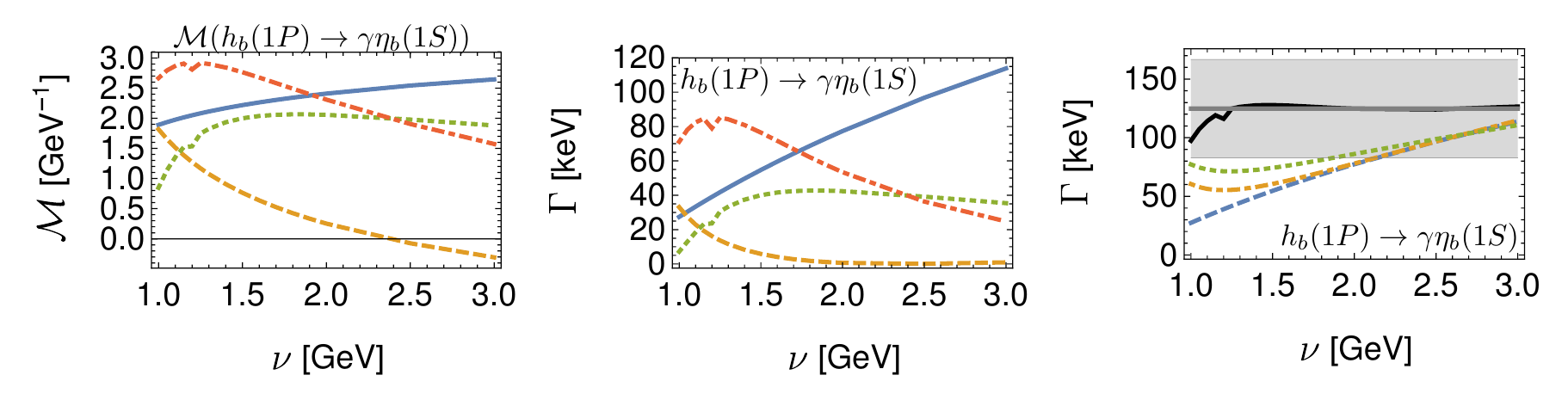}
\caption{\label{fig:21P111S0-orderBYorderFinalResult} Sum of the matrix 
elements shown in Figs.~\ref{fig:23PJ13S1-static} 
and~\ref{fig:21P111S0-others}, and decay width at different orders; also final 
decay width according to Eq.~\eqref{eq:FullDecayWidth2}, for the E1 transition 
$h_b(1P)\to \gamma\eta_b(1S)$. \textit{Left panel:} Sum of the matrix elements 
at (not up to) LO (solid blue), NLO (dashed orange), NNLO (dotted green) and 
NLO+NNLO (dot-dashed red). \textit{Middle panel:} Decay width at (not up to) LO 
(solid blue), NLO (dashed orange), NNLO (dotted green) and NLO+NNLO (dot-dashed 
red). \textit{Right panel:} Description as in 
Fig.~\ref{fig:23PJ13S1-FinalResult}.}
\end{figure*}

\subsubsection{\texorpdfstring{$h_b(1P)\to \gamma\eta_b(1S)$}{hb(1P) to gamma etab(1S)}}
We apply now the former analysis to the electric dipole transition $h_b(1P)\to 
\gamma\eta_b(1S)$. Figure~\ref{fig:21P111S0-relativistic} shows the LO decay 
rate and its correction due to higher order electromagnetic operators in the 
pNRQCD Lagrangian. In comparison with the $\chi_{bJ}$ radiative transitions, 
the LO transition width and the correction induced by higher order operators 
are larger in this case. This is because the photon energy, $k_\gamma$, is 
larger for increasing $J$ in the $\chi_{bJ}(1P))\to \gamma\Upsilon(1S)$ 
transitions and even larger in the $h_b(1P)\to \gamma\eta_b(1S)$ transition, 
see Eq.~\eqref{kgammavalues}. The fact that the photon energy enters with the 
third power in the expression of the decay width explains then the overall 
increasing effect. The correction due to higher order operators is about~$10\%$.

The corrections to the decay width due to the radiative corrections to the 
static potential \eqref{eq:StatPot} are the same as the ones shown in 
Fig.~\ref{fig:23PJ13S1-static}. As already mentioned, this is so because none 
of these potentials depends on either the spin $s$ or the total angular 
momentum $J$.

Figure~\ref{fig:21P111S0-others} shows the corrections to the matrix element of 
the LO electric dipole operator \eqref{LagE1}, due to the relativistic 
corrections to the bottomonium wave functions discussed in 
Sec.~\ref{subsec:RelativisticCorrectionspot}. These corrections contribute to 
the term $R_{21}^{S=0}$ in Eq.~\eqref{eq:FullDecayWidth2}. Since the initial 
and final states in the transition are now spin-singlet states, corrections to 
the wave functions due to the spin-orbit, spin-spin, and tensor potentials are 
absent. This has a major impact on the total NNLO matrix element because the 
correction induced by the $V_{S^2}^{(2)}$ potential is zero now, whereas in the 
$\chi_{bJ}$ case it is large (and negative) especially in the low $\nu$ region.

The left (middle) panel of Fig.~\ref{fig:21P111S0-orderBYorderFinalResult} 
shows for each order the sum of all matrix elements (decay widths) including 
radiative corrections to the static potential (see 
Fig.~\ref{fig:23PJ13S1-static}) and higher order relativistic corrections to 
the potential and kinetic energy (see Fig.~\ref{fig:21P111S0-others}). The 
kink, visible in the NNLO and thus in the NLO+NNLO matrix elements at about 
$1.2\,\text{GeV}$, can be traced back to the zero crossing or maximum in the 
matrix elements that account for the radiative corrections to the static 
potential. The absence of several negative contributions at NNLO yields a 
stronger dependence on the scale $\nu$ for values $\nu \lesssim 1.5\,\text{  
GeV}$ than in the $\chi_{bJ}$ case. In this region of $\nu$, the NLO+NNLO 
matrix element and the subsequent decay width clearly exceed the leading order 
ones.

The result that follows from summing up all previous corrections, \textit{i.e.},
those that contribute to the term $R^{S=0}_{21}$ in \eqref{eq:FullDecayWidth2},
shown in the first two panels of 
Fig.~\ref{fig:21P111S0-orderBYorderFinalResult}, and those that contribute to 
the terms other than $R^{S=0}_{21}$ in \eqref{eq:FullDecayWidth2}, shown in 
Fig.~\ref{fig:21P111S0-relativistic}, is shown in the right panel of 
Fig.~\ref{fig:21P111S0-orderBYorderFinalResult}. The renormalization scale 
dependence of the decay width is reduced when the NLO and NNLO corrections are 
included: by varying the renormalization scale from 1~GeV to 3~GeV the LO decay 
width spans over the range $(27$-$114)\,\text{keV}$, incorporating the NLO 
contribution shrinks the range to $(64$-$115)\,\text{keV}$, and incorporating 
the NNLO correction shrinks further the range to $(97$-$127)\,\text{ keV}$. 
This comes at the cost of an even worse convergence pattern of the perturbative 
series than in the $\chi_{bJ}$ case. We observe again a slight shift towards 
higher upper bounds, but the whole range and thus the overall scale dependence 
decreases.

Omitting the corrections to the decay width induced by the radiative 
corrections to the static potential results in a curve (green-dotted curve in 
the right panel of Fig.~\ref{fig:21P111S0-orderBYorderFinalResult}) that is 
quite close to the LO one at large values of $\nu$ and whose $\nu$-scale 
dependence is weaker than the complete result at low values of $\nu$. This is 
in contrast with the effect observed for the $\chi_{bJ}$ states, but 
understandable since several additional contributions appear in the $\chi_{bJ}$ 
case that are not present here.

We choose to set the central value of the decay width at $\nu = 
1.25\,\text{GeV}$, following the same prescription discussed for the 
$\chi_{bJ}$ case. The main differences in comparison to the $\chi_{bJ}$ 
transition width curves in Fig.~\ref{fig:23PJ13S1-FinalResult} are the overall 
weaker scale dependence, the worse convergence of the perturbative series, and 
the shape of the curve for large values of the renormalization scale $\nu$.
Assigning the error to the transition width as in the $\chi_{bJ} \to 
\gamma\Upsilon(1S)$ case discussed above, a fixed order determination at NNLO 
gives for the E1 transition width of the $h_b$:
\begin{equation}
\Gamma(h_b(1P)\to \gamma\eta_b(1S)) = 124^{+42}_{-42}~\text{keV} \,.
\end{equation}
In the following Sec.~\ref{sec:theory2}, we will see how to improve also this 
determination by resumming the known terms of the perturbative expansion of the 
static potential into the wave function.

\section{Numerical analysis in pNRQCD at weak coupling with resummation of the static potential}
\label{sec:theory2}

\subsection{Log resummation and renormalon subtraction}
We have seen in the previous section that the electric dipole transitions from 
the lowest-lying $P$-wave bottomonium states are not reliably described by 
fixed order calculations. The reason is that, even if these states are weakly 
coupled and the potential computable in perturbation theory, considering them 
at LO as Coulombic bound states is inadequate. Indeed, expanding around the 
Coulomb potential, $V_{s}^{(0)}$, has led to a poor convergence of the 
perturbative series, resulting in a strong dependence on the renormalization 
scale and large theoretical uncertainties.

We deal with this problem by rearranging the perturbative expansion in pNRQCD 
in such a way that the static potential is exactly included in the LO 
Hamiltonian. One motivation for this reorganization of the perturbative series 
is the observation, originating from Ref.~\cite{Pineda:2002se} (for more recent 
studies see, for instance, \cite{Bazavov:2014soa}), that, when comparing the 
static potential with lattice perturbation theory at short distances, the 
inclusion of higher order corrections is necessary to get a good agreement. An 
accurate treatment of the potential is particularly important for those 
observables, like the electric dipole transition widths, that are sensitive to 
the precise form of the wave function.

The new expansion scheme was applied in Ref.~\cite{Kiyo:2010jm} to study 
electromagnetic decays of heavy quarkonium, and in Ref.~\cite{Pineda:2013lta} 
to compute magnetic dipole transitions between low-lying heavy quarkonia. The 
effect of the new rearrangement was found to be large. In particular, the exact 
treatment of the soft logarithms of the static potential made the 
renormalization scale dependence much weaker. We proceed herein to apply the 
same scheme to the electric dipole transitions under study. Like in the 
magnetic dipole transition computation performed in Ref.~\cite{Pineda:2013lta}, 
an improvement in the convergence of the perturbative expansion is expected. 
The perturbative expansion will consist of just two terms: A leading order 
term, incorporating exactly the static potential, and a term incorporating the 
remaining corrections coming from higher order electromagnetic operators, and 
higher order relativistic corrections to the wave functions.

We follow the same setup of Ref.~\cite{Pineda:2013lta}. The leading order 
Hamiltonian reads now:
\begin{equation}
H_{\text{exact}\,V_s}^{(0)} = -\frac{\vec{\nabla}^2}{m} + V_{s}(r) \,,
\label{Hexact}
\end{equation}
where the static potential is ideally summed to all orders in perturbation 
theory. In practice, it is only known up to order $\alpha_{\textrm{s}}^{4}$, 
hence we take\footnote{To keep the notation simple, we will not explicitly 
write the dependence on the scale for quantities where this is due only to the 
truncation of the perturbative expansion.}
\begin{equation}
V_{s}(\nu_{\text{us}},r) = V_{s}^{(0)}(r) \left[ 1 + \sum\limits_{k=1}^{3} 
\left(\frac{\alpha_{\textrm{s}}}{4\pi}\right)^k 
a_k(\nu,\nu_{\text{us}},r) \right] \,.
\label{eq:StatPot2}
\end{equation}
The analytical expressions of $a_{1}(\nu,r)$ and $a_{2}(\nu,r)$ have been given 
in Eqs.~\eqref{eq:radiative1} and~\eqref{eq:radiative2}, respectively. The term 
$a_{3}(\nu,\nu_{\text{us}},r)$ is known from 
Refs.~\cite{Anzai:2009tm,Smirnov:2009fh}:
\begin{align}
a_3(\nu,\nu_{\text{us}},r) &= a_3+ a_1\beta_0^{\,2}\pi^2 + 
\frac{5\pi^2}{6}\beta_0\beta_1 + 16\zeta_3\beta_0^{\,3} \nonumber \\
& \hspace{-1.5cm}
+ \bigg(2\pi^2\beta_0^{\,3} + 6a_2\beta_0 + 4a_1\beta_1 + 2\beta_2 
+ 144\pi^2\bigg)\, \text{ln}\left(\nu e^{\gamma_E} r\right) \nonumber \\
& \hspace{-1.5cm}
+ \bigg(12a_1\beta_0^{\,2} + 10\beta_0\beta_1\bigg) \, \text{ln}^{2}\left(\nu 
e^{\gamma_E} r\right)\, + 8\beta_0^{\,3} \,\text{ln}^3\left(\nu e^{\gamma_E} 
r\right) \nonumber \\
& \hspace{-1.5cm}
+ \delta a_3^{\text{us}}(\nu,\nu_{\text{us}}) \,,
\end{align}
where $ \beta_2= 2857/2 - 5033 n_f /18 + 325 n_f^2/54$, $a_3$ may be read from 
the original literature or, for instance, from~\cite{Kiyo:2014uca}, and $\delta 
a_3^{\text{us}}(\nu,\nu_{\text{us}})$, encoding the subtraction of ultrasoft 
corrections from the static potential, is taken~\cite{Brambilla:1999qa}
\begin{equation}
\delta a_{3}^\text{us}(\nu,\nu_\text{us}) = 144 \pi^{2} \, 
\ln{\frac{\nu_\text{us}}{\nu}} \,.
\label{ultrasoft}
\end{equation}
Ultrasoft corrections to the static potential are due to gluons carrying energy 
and momentum of order $\alpha_{\textrm{s}}/r$; the scale $\nu_\text{us}$ is the 
factorization scale separating the ultrasoft energy and momentum region from 
higher ones. We will not resum here ultrasoft logs, like the one appearing in 
\eqref{ultrasoft}, although the result is known at 
leading~\cite{Pineda:2000gza} and next-to-leading 
accuracy~\cite{Brambilla:2009bi}. The reason is that their numerical effect is 
small with respect to other sources of error.

The perturbative expansion \eqref{eq:StatPot2} does not converge due to 
factorially growing terms that, once Borel resummed, give rise to singularities 
in the Borel plane, known as renormalons. The leading order renormalon 
affecting the static potential, $V_s$, cancels against twice the pole mass 
$m$~\cite{Pineda:1998id,Hoang:1998nz,Beneke:1998rk}. To make this cancellation 
explicit one adds/subtracts the same renormalon contribution from twice the 
pole mass/the static potential ensuring that both are expressed in series of 
$\alpha_{\textrm{s}}$ to the same power and at the same scale, \textit{e.g.}, 
$\nu$:
\begin{equation}
\begin{aligned}
m &= m_X+\delta m_X \,, \\
V_{s}(r) &= V_{s,X}(r)-2\, \delta m_X \,,
\end{aligned}
\label{VsRen}
\end{equation}
where
$\displaystyle 
\delta m_{X} = \nu_f \sum_{k=0}^{3} \delta m^{(k)}_{X} \! 
\left(\frac{\nu_f}{\nu}\right) \, \alpha_{\textrm{s}}^{k+1}(\nu)$
encodes the pole mass renormalon contribution, $\nu_f$ is the renormalon 
factorization scale and $X$ stands for the chosen renormalon subtraction scheme.
For the renormalon subtraction scheme we use here the RS$'$ 
scheme~\cite{Pineda:2001zq},\footnote{We have checked against the 
RS~\cite{Pineda:2001zq} and the potential subtracted (PS)~\cite{Beneke:1998rk} 
schemes that the LO matrix element depends only mildly on the adopted 
renormalon subtraction scheme.} which amounts at choosing
\begin{widetext}
\begin{align}
&
\delta m^{(0)}_{\text{RS}'}=0 \,,
\qquad\qquad\qquad
\delta m^{(1)}_{\text{RS}'}\!\left(\frac{\nu_f}{\nu}\right)= N_m 
\frac{\beta_{0}}{2\pi} S(1,b) \,, 
\label{RSini}\\
&
\delta m^{(2)}_{\text{RS}'}\!\left(\frac{\nu_f}{\nu}\right)= 
N_m \left(\frac{\beta_{0}}{2\pi}\right) \left[ S(1,b) 
\frac{2d_{0}(\nu,\nu_{f})}{\pi} + \left(\frac{\beta_{0}}{2\pi}\right) 
S(2,b)\right] \,, \\
&
\delta m^{(3)}_{\text{RS}'}\!\left(\frac{\nu_f}{\nu}\right) = N_m 
\left(\frac{\beta_{0}}{2\pi}\right) \left[ S(1,b) 
\frac{3d_{0}^{2}(\nu,\nu_{f})+2d_{1}(\nu,\nu_{f})}{\pi^{2}} + 
\left(\frac{\beta_{0}}{2\pi}\right) S(2,b) \frac{3d_{0}(\nu,\nu_{f})}{\pi} +
\left(\frac{\beta_{0}}{2\pi}\right)^{2} S(3,b) \right] \,,
\end{align}
where
\begin{equation}
S(n,b) = \sum_{k=0}^2 c_k \frac{\Gamma(n+1+b-k)}{\Gamma(1+b-k)} \,,
\qquad
d_k(\nu,\nu_f) = \frac{\beta_k}{2^{1+2k}}\ln{\frac{\nu}{\nu_f}} \,,
\end{equation}
with $b = {\beta_1}/{(2\beta_0^2)}$ and
\begin{equation}
c_0=1 \,, 
\qquad
c_1 = \frac{\beta_1^2-\beta_0\beta_2}{4\beta_0^4b} \,,
\qquad 
c_2 = \frac{\beta_1^4 + 4 \beta_0^3 \beta_1 \beta_2 - 2 \beta_0 \beta_1^2 
\beta_2 + \beta_0^2 (-2 \beta_1^3 + \beta_2^2) - 2 \beta_0^4 \beta_3}{32 
\beta_0^8 b(b-1)} \,.
\label{RSfin}
\end{equation}
\end{widetext}
The mass that we are using for the bottom quark is 
$m_{b,{\text{RS}'}}(\nu_{f}=1.0\,\text{GeV})=4.859\,\text{GeV}$. It can be 
translated into the $\overline{\text{MS}}$-mass: 
$m_{b}(m_{b})=4.19\,\text{GeV}$~\cite{Pineda:2006gx}. Our reference value for 
$N_m$ is $N_m=0.574974$ (for three light flavors) from 
Ref.~\cite{Pineda:2001zq}.\footnote{In the literature, there is an updated 
value, $N_{m}=0.563126$, from Ref.~\cite{Ayala:2014yxa}, as well as other 
recent determinations, like $N_{m}=0.535\pm 0.010$ from 
Ref.~\cite{Komijani:2017vep}. Since we have verified that these different 
determinations vary our results well inside the final errors, we will neglect 
in the following the uncertainty of~$N_m$.} As in the previous section, our 
reference value for $\alpha_{\textrm{s}}$ is 
$\alpha_{\textrm{s}}^{(n_{f}=3)}(1\,\text{GeV})=0.480$, and, like there, the 
running is implemented with four-loop accuracy. We set $\nu_{\text{us}}=\nu_f$. 
This choice is motivated by the fact that $\nu_{\text{us}}$ has to be smaller 
than the typical momentum transfer scale, i.e., $\nu_{\text{us}} < p \sim 1/a = 
1.25$~GeV on the one hand, and $\nu_{\text{us}}$ has to be larger than the 
scale where perturbation theory breaks down, say $0.7$~GeV. Varying 
$\nu_{\text{us}}$ from $0.7$~GeV to $1.25$~GeV induces a change from $+4$\% to 
$-2$\% in the coefficient $\delta a_{3}^\text{us}(\nu,\nu_\text{us})$. The 
numerical impact of this change in the three loop coefficient of the static 
potential is negligible with respect to the dependence on the scale $\nu$. This 
is not surprising as ultrasoft corrections are beyond the accuracy of the 
present study.

\begin{figure*}[!t]
\centering
\includegraphics[width=0.98\textwidth]{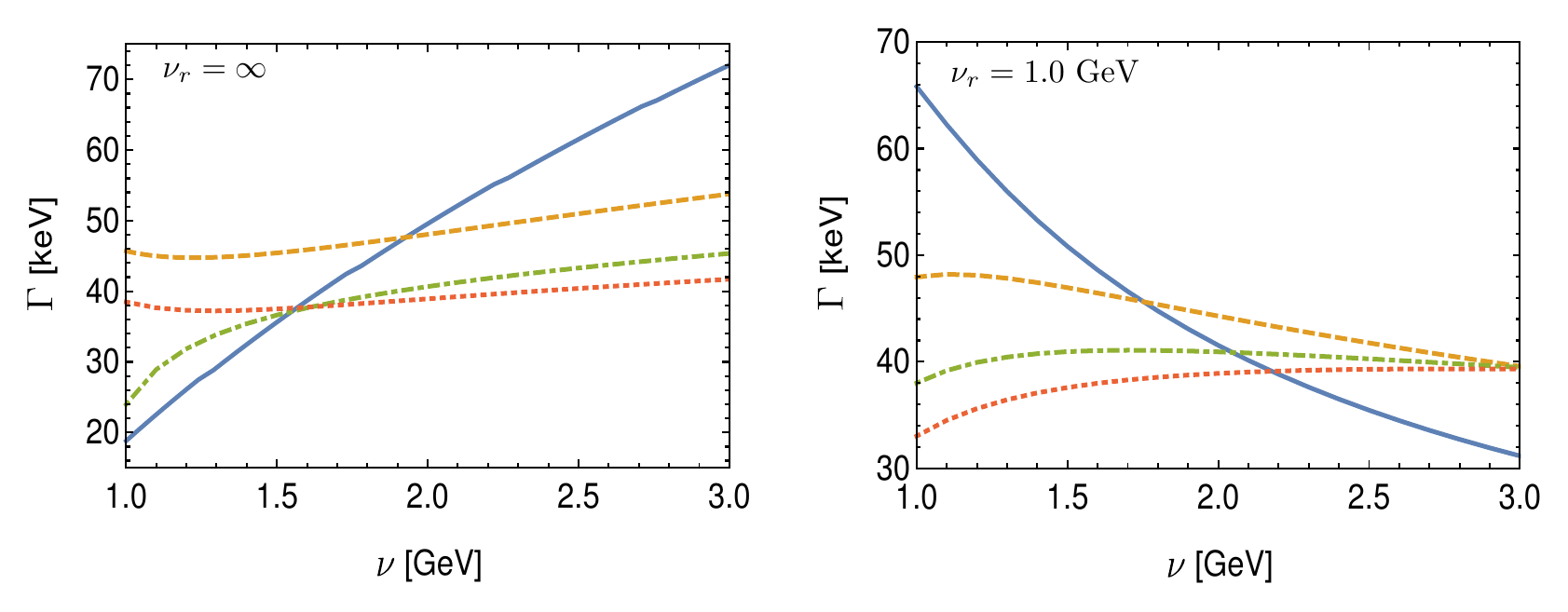}
\caption{\label{fig:Renormalon+RGI} Leading order decay rate of the electric 
dipole transition $\chi_{b1}\to \gamma\Upsilon(1S)$. The renormalon subtracted 
static potential is included at different orders in the Schr\"odinger equation, 
which is solved exactly by numerical methods: LO Coulomb-like potential (solid 
blue), potential up to NLO (dashed orange), potential up to NNLO (dot-dashed 
green) and potential up to NNNLO (dotted red). The left panel shows the case 
where the coupling in the static potential is computed at the fixed scale $\nu$ 
(case $\nu_{r} = \infty$ in Eq.~\eqref{eq:RSpScheme}), while the right panel 
shows the case where the coupling in the static potential is computed at the 
scale $1/r$ at short distances (case $\nu_{r} = 1.0~\text{GeV}$ in 
Eq.~\eqref{eq:RSpScheme}).}
\end{figure*}

In the short range, it is possible to further improve the static potential by 
resumming potentially large logs of the type $\ln{\nu r}$ by setting the scale 
$\nu = 1/r$ and yet achieve renormalon cancellation order by order in 
$\alpha_{\textrm{s}}(1/r)$ (see~\cite{Pineda:2002se}). 
Following~\cite{Pineda:2013lta}, we finally define our renormalon subtracted 
static potential in the RS$'$ scheme as
\begin{widetext}
\begin{equation}
\label{eq:RSpScheme}
V_{s,\text{RS}'}(\nu,\nu_f,\nu_r,r) =
\begin{cases}
\displaystyle V_{s} + 2 \delta m_{\text{RS}'} \big|_{\nu=1/r} \equiv 
\sum_{k=0}^{3} V_{s,\text{RS}'}^{(k)} \alpha_{\textrm{s}}^{k+1}(1/r) & \mbox{if 
} r<\nu_{r}^{-1} \,, \\
\displaystyle V_{s} + 2 \delta m_{\text{RS}'} \big|_{\nu=\nu} \equiv 
\sum_{k=0}^{3} V_{s,{\text{RS}'}}^{(k)} \alpha_{\textrm{s}}^{k+1}(\nu) & 
\mbox{if } r>\nu_{r}^{-1} \,. \\
\end{cases}
\end{equation}
\end{widetext}
The scale $\nu_r$ separates short distances, where logs are resummed in the 
coupling  ($r < 1/\nu_r$), from long distances, where the coupling is evaluated 
at the fixed scale $\nu$  ($r > 1/\nu_r$). If $\nu_r = \infty$, this is 
equivalent to compute with a fixed scale over all distances; if $\nu_r = 0$, 
this is equivalent to compute the coupling at $1/r$ over the full distance 
range.

The renormalon factorization scale, $\nu_f$, must be chosen low enough that the 
subtracted mass, $\delta m_{\text{RS}'}$, does not jeopardize the power 
counting, \textit{i.e.}, $\delta m_{\text{RS}'}$ must be of order $mv^2$ or 
smaller, but also large enough that $\delta m_{\text{RS}'}$ encompasses the 
renormalon, \textit{i.e.}, the renormalon subtracted series converges, and 
perturbation theory holds. In our analysis, we observe that we can use the 
rather low value $\nu_f=1.0\,\text{GeV}$ and yet achieve renormalon 
cancellation. Other choices of $\nu_f$ are possible, but, given the above 
constraints, the allowed range of variation for $\nu_f$ is even more restricted 
than for $\nu_{\text{us}}$. In Refs.~\cite{Pineda:2013lta,Peset:2018jkf} the 
effect of taking $\nu_f = 0.7$~GeV has been considered. The impact on the 
bottomonium mass is at most $1\%$. We consider this to be a reasonable upper 
limit also for the transition widths. The uncertainty coming from the scale 
$\nu_f$ (as well as the one from the scale $\nu_{\text{us}}$ considered before)
is, therefore, negligible with respect to the one coming from the scale $\nu$, 
which is, on the overall, the largest theoretical uncertainty in our 
computation.

We can look at the effects on the leading order transition width, 
\textit{i.e.}, the matrix element of the leading order E1 operator 
\eqref{LagE1}, when incorporating the static potential \eqref{eq:RSpScheme} at 
different perturbative orders into the exact solution of the Schr\"odinger 
equation. Differently from the previous section, now the Schr\"odinger equation 
with the potential \eqref{eq:RSpScheme} can be solved, beyond LO, only 
numerically. We provide some details on the numerical solution of the 
Schr\"odinger equation in Appendix~\ref{app:SolvingSchroedingerEquation}.

Let us consider, as an example, the transition $\chi_{b1}(1P) \to 
\gamma\Upsilon(1S)$; the other transitions at leading order follow from this 
one just by rescaling all the curves by the constant factor 
$(k_\gamma/423\,\text{MeV})^3$, which corrects for the photon energy. The left 
panel of Fig.~\ref{fig:Renormalon+RGI} shows the leading order transition rate 
when the coupling in the static potential is computed at the fixed scale $\nu$, 
corresponding to the case $\nu_{r} = \infty$ in Eq.~\eqref{eq:RSpScheme}. 
Solving the Schr\"odinger equation with only the Coulomb-like term in the 
static potential gives back the same LO result as in Sec.~\ref{sec:theory1}. 
This decay rate (solid blue curve) depends strongly on the renormalization 
scale: It ranges from $18\,\text{keV}$ to $72\,\text{keV}$ when running $\nu$ 
from $1\,\text{GeV}$ to $3\,\text{GeV}$. However, the $\nu$-scale dependence 
becomes mild as NLO (dashed orange curve), NNLO (dot-dashed green curve) and 
NNNLO (dotted red curve) radiative corrections to the static potential are 
added to the Schr\"odinger equation. Indeed, the decay rate changes only of 
about $4\,\text{keV}$ over the considered $\nu$-range, when the three loop 
static potential is considered. Moreover, the convergence of the perturbative 
series has improved with respect to the fixed order case. Convergence tends to 
worsen only for low $\nu$.

The right panel of Fig.~\ref{fig:Renormalon+RGI} shows the same quantity when 
the coupling in the static potential is computed at the scale $1/r$ for $r< 
1.0~\text{GeV}^{-1}$ and at the scale $\nu$ for $r> 1.0~\text{GeV}^{-1}$, 
corresponding to the case  $\nu_{r} = 1.0~\text{GeV}$ in 
Eq.~\eqref{eq:RSpScheme}. The perturbative series appears to converge  over the 
whole range $1\,\text{GeV} \le \nu \le 3\,\text{GeV}$, and in particular for 
low $\nu$. As for the curves in the left panel, also for the curves shown in 
the right panel the dependence on the renormalization scale becomes mild with 
increasing order: At NNNLO the decay rate changes by less than $8\,\text{keV}$ 
when $\nu$ goes from $1$ to $3\,\text{GeV}$, which is slightly more than for 
the corresponding NNNLO decay rate in the left panel.

\begin{figure}[!t]
\centering
\includegraphics[width=0.44\textwidth]{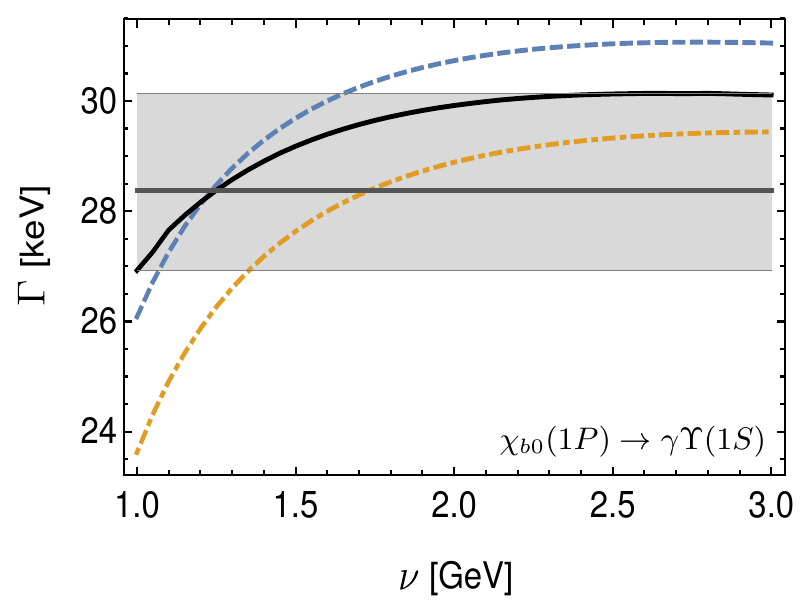}
\caption{\label{fig:GfinalChib0} Decay width, according to 
Eq.~\eqref{eq:FullDecayWidth}, for the transition $\chi_{b0}(1P)\to 
\gamma\Upsilon(1S)$ in the scheme discussed in the text. The dashed blue curve 
is the leading order decay rate, the dot-dashed orange curve includes 
contributions stemming from higher order electromagnetic operators in the 
pNRQCD Lagrangian and the solid black curve is the final result including both  
contributions from higher order electromagnetic operators and relativistic 
corrections to the wave functions of the initial and final states. We take our 
final value at $\nu = 1.25\,\text{GeV}$ and the gray band indicates the 
associated uncertainty.}
\end{figure}

\begin{figure}[!t]
\centering
\includegraphics[width=0.44\textwidth]{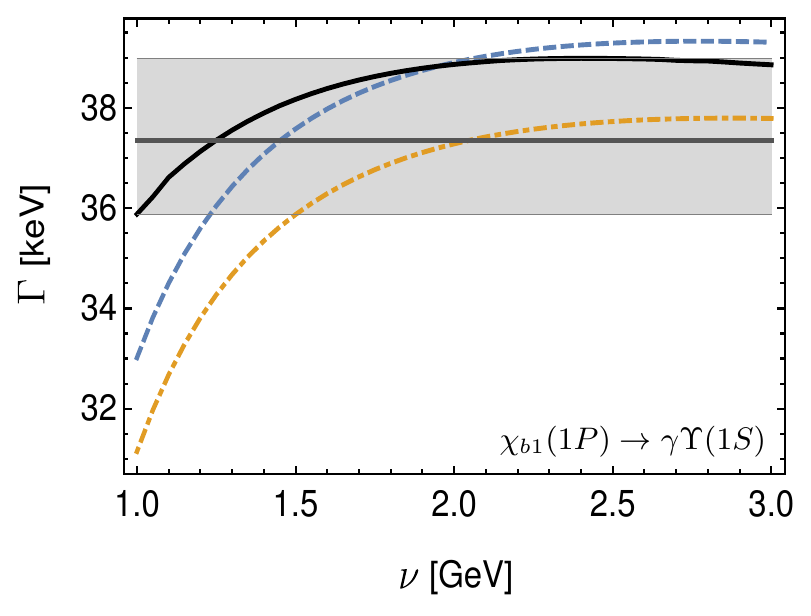}
\caption{\label{fig:GfinalChib1} Decay width, according to 
Eq.~\eqref{eq:FullDecayWidth}, for the transition $\chi_{b1}(1P)\to 
\gamma\Upsilon(1S)$. Description is as in Fig.~\ref{fig:GfinalChib0}.}
\end{figure}

\subsection{Numerical analysis}
We are now in the position to discuss the final determinations of the electric 
dipole transitions $\chi_{bJ}(1P)\to \gamma\Upsilon(1S)$ with $J=0,\,1,\,2$ and 
$h_{b}(1P)\to \gamma\eta_{b}(1S)$. We use wave functions obtained from the 
solution of the Schr\"odinger equation with the full static potential 
\eqref{eq:RSpScheme}. The static potential is taken at three loops, 
Eq.~\eqref{eq:StatPot2}, including ultrasoft effects. The leading order 
renormalon is subtracted according to the RS$'$ scheme defined in 
Eqs.~\eqref{RSini}-\eqref{RSfin}. The relevant factorization scales are set to 
be $\nu_{\text{us}} = \nu_{r} = \nu_{f} = 1.0~\text{GeV}$. Higher order 
corrections of relative order $v^2$ come from higher order electromagnetic 
operators in the pNRQCD Lagrangian, terms in \eqref{eq:FullDecayWidth} and 
\eqref{eq:FullDecayWidth2} other than $R^{S=1}_{21}(J)$  and $R^{S=0}_{21}$, 
respectively, and from higher order relativistic corrections affecting initial 
and final states, terms contributing to $R^{S=1}_{21}(J)$  and $R^{S=0}_{21}$ 
in \eqref{eq:FullDecayWidth} and \eqref{eq:FullDecayWidth2}, respectively, and 
stemming from Eq.~\eqref{eq:deltaH}.

\subsubsection{\texorpdfstring{$\chi_{bJ}(1P)\to \gamma\Upsilon(1S)$ with $J=0,1,2$}{chibJ(1P)to gamma Upsilon(1S) with J=0,1,2}}
The decay width for the $\chi_{b0}(1P)\to \gamma\Upsilon(1S)$ transition is 
shown in Fig.~\ref{fig:GfinalChib0}. The leading order (full $V_{s}$) 
non-relativistic decay rate is the dashed blue curve, the dot-dashed orange 
curve includes relativistic contributions stemming from higher order 
electromagnetic operators and the solid black one includes both  contributions 
from higher order electromagnetic operators and relativistic corrections to the 
wave functions of the initial and final states.

The leading order decay width depends weakly on the renormalization scale: It 
varies from $\Gamma\approx 26\,\text{keV}$ at $\nu=1\,\text{GeV}$ to 
$\Gamma\approx 31\,\text{keV}$ at $\nu=3\,\text{GeV}$. This feature is 
preserved when higher order electromagnetic operators are included and also in 
the final result. In fact, the $\nu$-dependence of the final result, which is 
about $3\,\text{keV}$, is weaker than that of the leading order result and also 
weaker than that obtained from including only higher order electromagnetic 
operators. A variation of $3\,\text{keV}$ over a central value of about 
$28\,\text{keV}$ represents an uncertainty of about $11\%$ in our determination 
of the decay rate. Moreover, higher order electromagnetic operators and 
relativistic corrections to the initial and final states provide relatively 
small changes to the LO transition width.

The gray error band accounts for the uncertainty due to the unknown higher 
order terms in the perturbative expansion. This is computed, here and in the 
following plots, by taking the largest between the variation of the result with 
the scale and one half of the maximum difference between the leading order and 
the final result, as described in the previous section after 
Eq.~\eqref{Bohrradius}.

An interesting feature of Fig.~\ref{fig:GfinalChib0} is that the corrections 
induced by higher order electromagnetic operators diminish the LO decay rate, 
whereas relativistic corrections to the initial and final states increase it. 
As a result, at the renormalization scale $\nu = 1.25$~GeV, the value of the 
decay width $\Gamma(\chi_{b0}(1P)\to \gamma\Upsilon(1S))$ turns out to be very 
similar to the LO result. This will not be the case for the other transitions.

\begin{figure}[!t]
\centering
\includegraphics[width=0.44\textwidth]{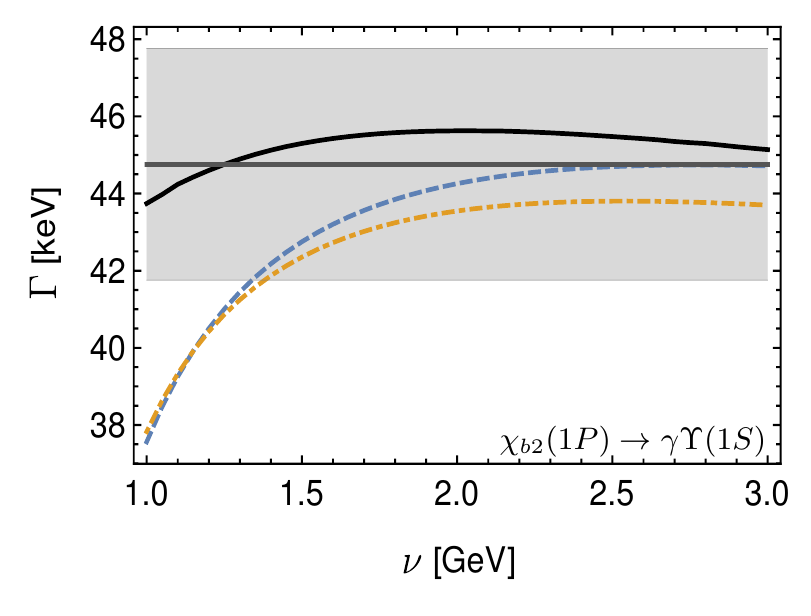}
\caption{\label{fig:GfinalChib2} Decay width, according to 
Eq.~\eqref{eq:FullDecayWidth}, for the transition $\chi_{b2}(1P)\to 
\gamma\Upsilon(1S)$. Description is as in Fig.~\ref{fig:GfinalChib0}.}
\end{figure}

We have performed the same analysis for the electric dipole transitions 
$\chi_{b1}(1P)\to \gamma\Upsilon(1S)$ and $\chi_{b2}(1P)\to 
\gamma\Upsilon(1S)$ in Figs.~\ref{fig:GfinalChib1} and~\ref{fig:GfinalChib2}, 
respectively. Similar features, as the one observed in the $\chi_{b0}(1P)\to 
\gamma\Upsilon(1S)$ case, are seen here, too. However, we notice that the 
effect due to relativistic corrections to the initial and final states is a 
factor $2$-$3$ larger in these cases. We also observe that the final decay 
rates for $\chi_{b1}(1P)\to \gamma\Upsilon(1S)$ and $\chi_{b2}(1P)\to 
\gamma\Upsilon(1S)$ show a weaker dependence on the renormalization scale than 
for $\chi_{b0}(1P)\to \gamma\Upsilon(1S)$. The scale variation for
$\chi_{b1}(1P)\to \gamma\Upsilon(1S)$ is $\lesssim 8\%$, and the scale 
variation for $\chi_{b2}(1P)\to \gamma\Upsilon(1S)$ is $\lesssim 5\%$. Finally, 
we remark that for the decay width $\Gamma(\chi_{b2}(1P)\to 
\gamma\Upsilon(1S))$, the LO result at the renormalization scale $\nu = 
1.25$~GeV is outside the final result error band.

\begin{figure}[!t]
\centering
\includegraphics[width=0.44\textwidth]{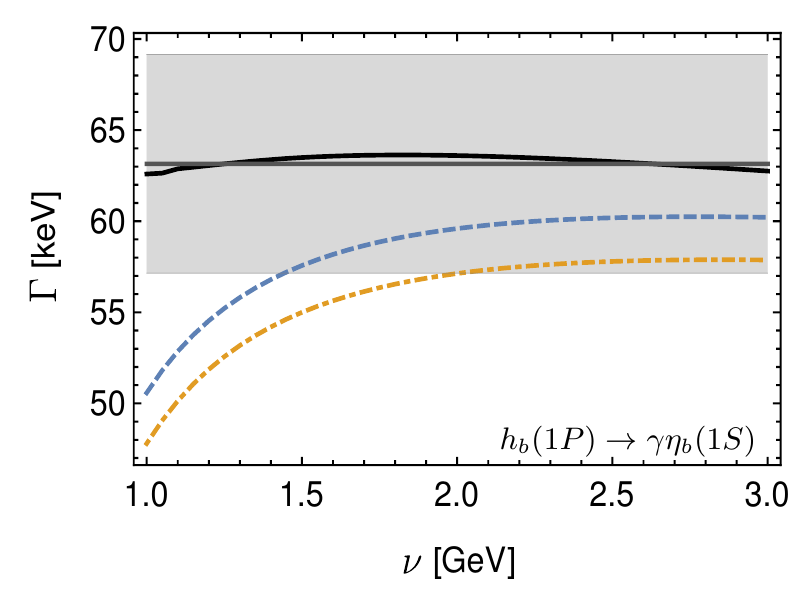}
\caption{\label{fig:Gfinalhb1} Decay width, according to 
Eq.~\eqref{eq:FullDecayWidth2}, for the transition $h_b(1P)\to 
\gamma\eta_b(1S)$. Description is as in Fig.~\ref{fig:GfinalChib0}.}
\end{figure}

\begin{table*}[!t]
\centering
\begin{tabular}{c|cc|cccccc}
\hline
\hline
Mode & LO & NNLO & CQM & R & GI & BT & LFQM & SNR$_{0/1}$ \\
\hline
$\chi_{b0}(1P)\to \gamma\Upsilon(1S)$ & $28.5$ & $28.4$ & $28.1$ & $29.9$ & 
$23.8$ & $25.7$ & - & $26.6 / 24.3$ \\
$\chi_{b1}(1P)\to \gamma\Upsilon(1S)$ & $36.0$ & $37.4$ & $35.7$ & $36.6$ & 
$29.5$ & $29.8$ & - & $33.6 / 30.0$ \\
$\chi_{b2}(1P)\to \gamma\Upsilon(1S)$ & $41.0$ & $44.8$ & $39.2$ & $40.2$ & 
$32.8$ & $33.0$ & - & $38.2 / 32.6 $ \\
$h_b(1P) \to \gamma\eta_b(1S)$ & $55.2$ & $63.2$ & $43.7$ & $52.6$ & $35.7$ & - 
& $37.5$ & $55.8 / 36.3$ \\
\hline
\hline
\end{tabular}
\caption{\label{tab:ResultsComparisonNum} The column LO lists our leading order 
results (blue dashed curves in Figs.~\ref{fig:GfinalChib0}-\ref{fig:Gfinalhb1}) 
and the column NNLO our final results (solid black curves in 
Figs.~\ref{fig:GfinalChib0}-\ref{fig:Gfinalhb1}), both taken at $\nu = 
1.25$~GeV. We compare them with those reported by a non-relativistic 
constituent quark model (CQM)~\cite{Segovia:2016xqb}, a relativistic quark 
model (R)~\cite{Ebert:2002pp}, a study based on the Godfrey--Isgur model 
(GI)~\cite{Godfrey:2015dia}, a study based on the Buchm\"uller--Tye potential 
(BT)~\cite{Grotch:1984gf}, a light-front quark model (LFQM)~\cite{Shi:2016cef}, 
and a screened potential model with zeroth-order wave functions (SNR$_0$) and 
first-order relativistically corrected wave functions 
(SNR$_1$)~\cite{Li:2009nr}. All decay widths are given in units of keV.}
\end{table*}

\subsubsection{\texorpdfstring{$h_b(1P)\to \gamma\eta_b(1S)$}{hb(1P) to gamma etab(1S)}}
Figure~\ref{fig:Gfinalhb1} shows the results for the $h_{b}(1P)\to 
\gamma\eta_b(1S)$ transition. The corrections to the decay width induced by 
higher order electromagnetic operators are very similar to the ones obtained in 
the previous cases. Their effect is to reduce the LO decay rate by about 
$2$-$3\,\text{keV}$. However, the effect due to relativistic corrections to the 
initial and final state wave functions is larger for the $h_{b}(1P)\to 
\gamma\eta_b(1S)$ transition than for the three transitions considered before. 
In particular, the decay width changes from about $52\,\text{keV}$ (dot-dashed 
orange curve) to about $63\,\text{keV}$ (solid black curve) at 
$\nu=1.25\,\text{GeV}$. This is because the initial and final state bottomonia 
in the transition are spin-singlet states and thus many corrections to the wave 
functions, like those induced by the spin-orbit, spin-spin and tensor 
potentials, are absent. In the case of the $\chi_{bJ}$ states, since they are 
spin-triplets, these corrections appear and tend to compensate other 
relativistic corrections due to different relative signs.

Similarly to the case of the $\chi_{b1}(1P)$ and $\chi_{b2}(1P)$ electric 
dipole transitions, also the decay width of the $h_{b}(1P)\to \gamma\eta_b(1S)$ 
transition displays a very weak dependence on $\nu$. The rate varies by a mere  
$1\,\text{keV}$ along the whole range of the renormalization scale studied 
herein. For this reason and for the one given in the paragraph above, the 
$h_{b}(1P)\to \gamma\eta_{b}(1S)$ decay width appears to be a well suited 
observable for studying relativistic corrections to the heavy quarkonium wave 
function. However, the uncertainty due to possible higher order corrections in 
the perturbative expansion, estimated by looking at one half of the maximum 
difference between the leading order and the final result, is about six times 
larger than the one coming from the scale variation in the transition. It is 
also larger than in the case of the $\chi_{bJ}$ transitions. This reflects in a 
larger final theoretical uncertainty. A related feature is that for the decay 
width $\Gamma(h_{b}(1P)\to \gamma\eta_b(1S))$, the LO result is outside the 
final result error band for $\nu = 1.25\,\text{GeV}$. As in the previous 
section, we choose to set the central value of the decay widths at the scale 
that self-consistently solves the Bohr-like radius equation \eqref{Bohrradius}. 
This scale is $\nu = 1/a = 1.25\,\text{GeV}$.

\subsection{Summary and comparisons}
Our final results for the electric dipole transitions $\chi_{bJ}(1P)\to 
\gamma\Upsilon(1S)$, with $J=0,\,1,\,2$, and $h_{b}(1P)\to \gamma\eta_{b}(1S)$, 
at relative order $v^2$ in the counting scheme adopted in this section that 
consists in treating the whole static potential as a leading order 
contribution, read
\begin{align}
\Gamma(\chi_{b0}(1P)\to \gamma\Upsilon(1S)) &= 28^{+2}_{-2}~\text{keV} \,, 
\label{finchi0}\\
\Gamma(\chi_{b1}(1P)\to \gamma\Upsilon(1S)) &= 37^{+2}_{-2}~\text{keV} \,, 
\label{finchi1}\\
\Gamma(\chi_{b2}(1P)\to \gamma\Upsilon(1S)) &= 45^{+3}_{-3}~\text{keV} \,, 
\label{finchi2}\\
\Gamma(h_b(1P)\to \gamma\eta_b(1S)) &= 63^{+6}_{-6}~\text{keV} \label{finh}\,.
\end{align}
Because of the very mild dependence on the renormalization scale, and the good 
convergence of the perturbative series, the results appear solid and their 
associated uncertainties are small. The uncertainties correspond to the gray 
bands shown in Figs.~\ref{fig:GfinalChib0}-\ref{fig:Gfinalhb1}, and have been 
computed as described after Eq.~\eqref{Bohrradius}. In the plots, the errors 
have not been rounded.

We compare our results with those obtained in several other theoretical 
approaches in Table~\ref{tab:ResultsComparisonNum}. These are a 
non-relativistic constituent quark model (CQM)~\cite{Segovia:2016xqb}, a 
relativistic quark model (R)~\cite{Ebert:2002pp}, a study based on the 
Godfrey--Isgur model (GI)~\cite{Godfrey:2015dia}, a study based on the 
Buchm\"uller--Tye potential model (BT)~\cite{Grotch:1984gf}, a light-front 
quark model (LFQM)~\cite{Shi:2016cef}, and a screened potential model with 
zeroth-order wave functions (SNR$_0$) and first-order relativistically 
corrected wave functions (SNR$_1$)~\cite{Li:2009nr}. 
Reference~\cite{Grotch:1984gf} does not provide a prediction for the 
$h_{b}(1P)\to \gamma\eta_{b}(1S)$ width, whereas Ref.~\cite{Shi:2016cef} is 
restricted to the study of the  $h_{b}(1P)\to \gamma\eta_{b}(1S)$ transition 
only. Our results agree well with those of other approaches for the 
$\chi_{b0}(1P)\to \gamma\Upsilon(1S)$ and $\chi_{b1}(1P)\to \gamma\Upsilon(1S)$ 
transitions. In the  case of the  $\chi_{b2}(1P)\to \gamma\Upsilon(1S)$ 
transition our result is slightly larger than the bulk of the other 
predictions, whereas in the case of the $h_{b}(1P)\to \gamma\eta_{b}(1S)$ 
transition it is significantly larger. The reasons for the differences may be 
diverse, and follow from the theoretical approaches listed in 
Table~\ref{tab:ResultsComparisonNum} being, to various degrees, 
phenomenological models that neither include QCD corrections in a systematic 
way, nor derive their parameters from QCD. Hence, they differ from our model 
independent determination in more than one way. For example, 
Refs.~\cite{Segovia:2016xqb,Godfrey:2015dia} do not include spin-independent 
$1/m$ and $1/m^2$ potentials, while Refs.~\cite{Grotch:1984gf,Li:2009nr} miss 
the $1/m$ potential.

\begin{table}[!t]
\centering
\scalebox{0.98}{\begin{tabular}{cccc}
\hline
\hline
Mode & $\mathcal{B}_i = \Gamma_i/\Gamma$ & $\Gamma_i$ & $\Gamma$ \\
\hline
$\chi_{b0}(1P)\to \gamma\Upsilon(1S)$ & $(1.94 \pm 0.27)\%$ & 
$28^{+2}_{-2}$~keV & $1.46^{+0.2}_{-0.2}$~MeV \\
$\chi_{b1}(1P)\to \gamma\Upsilon(1S)$ & $(35.0 \pm 2.1)\%$ & $37^{+2}_{-2}$~keV 
& $107^{+9}_{-9}$~keV \\
$\chi_{b2}(1P)\to \gamma\Upsilon(1S)$ & $(18.8 \pm 1.1)\%$ & $45^{+3}_{-3}$~keV 
& $238^{+21}_{-21}$~keV \\
$h_b(1P)\to \gamma\eta_b(1S)$ & $(52^{+6}_{-5})\%$ & $63^{+6}_{-6}$~keV & 
$121^{+18}_{-16}$~keV \\
\hline
\hline
\end{tabular}}
\caption{\label{tab:ResultsNum} Predicted total decay widths of the 
$\chi_{bJ}(1P)$ and  $h_{b}(1P)$ bottomonium states (column $\Gamma$), 
following our determinations of the bottomonium E1 transition widths (column 
$\Gamma_i$) and the experimental branching fractions reported by the PDG  
(column $\mathcal{B}_i = \Gamma_i/\Gamma$)~\cite{PhysRevD.98.030001}. The  
errors are obtained via standard Gaussian uncertainty propagation.}
\end{table}

Our final results \eqref{finchi0}-\eqref{finh} are predictions, as the 
bottomonium $P$-wave E1 transition widths have not been measured so far. In 
fact, for these electromagnetic transitions only the branching fractions are 
known, while there are no measurements of any of the total decay widths of the 
$\chi_{bJ}$, with $J=0,\,1,\,2$, and $h_{b}$ states. Nevertheless, we can use 
the branching fractions given by the PDG~\cite{PhysRevD.98.030001} and our 
results for the decay rates of the electric dipole transitions to predict the 
total decay widths of the $\chi_{bJ}(1P)$ and $h_{b}(1P)$ bottomonia. The 
results are given in Table~\ref{tab:ResultsNum}, where the errors are obtained 
via standard Gaussian uncertainty propagation. The Belle collaboration has 
reported an upper limit on the total decay width of the $\chi_{b0}(1P)$ at 
$90\%$ confidence level~\cite{Abdesselam:2016xbr}: 
$\Gamma(\chi_{b0}(1P))<2.4\,\text{MeV}$, which is compatible with our 
prediction.

\section{Conclusion}
\label{sec:conclusion}
We have computed the electric dipole transitions $\chi_{bJ}(1P)\to 
\gamma\Upsilon(1S)$, with $J=0,\,1,\,2$, and $h_{b}(1P)\to \gamma\eta_{b}(1S)$, 
within potential non-relativistic QCD, assuming that the typical binding energy 
scale, $mv^2$, is much larger than $\Lambda_{\text{QCD}}$, where $m$ is the 
mass of the heavy quark and $v$ its relative velocity. Consequences of this 
assumption are that $n=2$, $\ell =1$ bottomonia are taken as weakly-coupled 
bound states, and that non-perturbative effects are smaller than the accuracy 
reached in the calculation. This assumption would not be suited for  $n=2$, 
$\ell =1$ charmonia.

The precision that we have reached in this paper is $k_{\gamma}^{3}/(mv)^{2} 
\times \mathcal{O}(v^{2})$, $k_{\gamma}$ being the photon energy. At relative 
order $v^2$ we have included higher order electromagnetic interactions in the 
pNRQCD Lagrangian and higher order relativistic corrections to the initial and 
final state bottomonia, due to $1/m$ and $1/m^2$ potentials, and $1/m^3$ 
relativistic corrections to the kinetic energy. Concerning radiative 
corrections to the static potential, we have included them in two different 
counting schemes: in Sec.~\ref{sec:theory1}, perturbatively, counting higher 
order corrections as perturbations of the leading order Coulomb-like  
potential, and, in Sec.~\ref{sec:theory2}, non-perturbatively, counting all 
known terms in the perturbative expansion of the static potential as leading 
order and including them in the numerical solution of the Schr\"odinger 
equation for the initial and final state wave functions.

We summarize the main conclusions drawn from the first scheme. \textit{(i)} The 
decay widths show a strong dependence on the renormalization scale $\nu$. At 
leading order, the strong dependence is due to the running of 
$\alpha_{\textrm{s}}(\nu)$, which affects primarily the Bohr-like radius 
entering the initial and final state wave functions. At higher orders a 
significant $\nu$-dependence persists, due to the corrections to the initial 
and final state wave functions induced by the radiative corrections of the 
static potential. The static potential contains terms proportional to powers of 
$\ln{\nu r}$ that become large at low values of $\nu$. \textit{(ii)} Most of 
the corrections to the decay rates induced by the $1/m$ and $1/m^2$ potentials 
are relatively small and do not change much as a function of the 
renormalization scale $\nu$. The largest contributions come from the $1/m$ 
potential and the spin-spin one, especially for low values of the scale $\nu$. 
\textit{(iii)} The convergence of the perturbative series for all the studied 
electric dipole transitions is poor. This indicates that bottomonium $1P$ 
states are difficult to accommodate in this scheme. An observation that led us 
to adopt for our final analysis the second scheme.

In the second scheme, the Schr\"odinger equation is solved at leading order 
with all known terms of the perturbative static potential included, 
\textit{i.e.}, up to three loops. Further, we subtract to the static potential 
the leading order renormalon and resum at short distances potentially large 
logs of the type $\ln{\nu r}$. The main effects are: \textit{(i)} The leading 
order decay rates depend weakly on the renormalization scale and this is also 
so when higher order electromagnetic operators and relativistic corrections to 
the initial and final states are taken into account at relative order $v^2$. 
\textit{(ii)} Both $\mathcal{O}(v^2)$ corrections do not change much as 
functions of the renormalization scale and produce corrections to the leading 
order decay widths that are relatively small. \textit{(iii)} The corrections 
induced by higher order electromagnetic operators tend to diminish the leading 
order decay rates, whereas the opposite effect is found for the relativistic 
corrections to the initial and final state wave functions. These observations 
support our initial assumptions on the nature of the $1P$ bottomonia. Because 
the perturbative series appears convergent and only mildly dependent on the 
renormalization scale, the final results are affected by small uncertainties.

If the most critical of our assumptions, $mv^2 \gg \Lambda_{\text{QCD}}$, is 
relaxed to $mv^2 \sim \Lambda_{\text{QCD}}$, then non-perturbative corrections 
may become as large as the $\mathcal{O}(v^2)$ corrections considered above. 
Since the uncertainties on our final results have been chosen to include one 
half of the $\mathcal{O}(v^2)$ corrections, the effect of assuming  $mv^2 \sim 
\Lambda_{\text{QCD}}$ would be (at least) to double our final errors. A 
challenging alternative is to compute the non-perturbative contributions listed 
in Ref.~\cite{Brambilla:2012be}.

Finally, for ease of reference, we quote here again our final predictions for 
the $1P$ bottomonium electric dipole transitions:
\begin{align}
\Gamma(\chi_{b0}(1P)\to \gamma\Upsilon(1S)) &= 28^{+2}_{-2}~\text{keV} \,, \\
\Gamma(\chi_{b1}(1P)\to \gamma\Upsilon(1S)) &= 37^{+2}_{-2}~\text{keV} \,, \\
\Gamma(\chi_{b2}(1P)\to \gamma\Upsilon(1S)) &= 45^{+3}_{-3}~\text{keV} \,, \\
\Gamma(h_b(1P)\to \gamma\eta_b(1S)) &= 63^{+6}_{-6}~\text{keV} \,.
\end{align}
We have used the experimental branching fractions given by the PDG and the 
above theoretical results to predict the total decay widths of the 
$\chi_{bJ}(1P)$, with $J=0,\,1,\,2$, and $h_{b}(1P)$ bottomonia:
\begin{align}
\Gamma(\chi_{b0}(1P)) &= 1.46^{+0.2}_{-0.2}~\text{MeV} \,, \\
\Gamma(\chi_{b1}(1P)) &= 107^{+9}_{-9}~\text{keV} \,, \\
\Gamma(\chi_{b2}(1P)) &= 238^{+21}_{-21}~\text{keV} \,, \\
\Gamma(h_b(1P))       &= 121^{+18}_{-16}~\text{keV} \,.
\end{align}
These numbers could be of interest for future experimental determinations, for 
instance, at Belle II.

\begin{acknowledgments}
We thank Nora Brambilla, Yuichiro Kiyo, Clara Peset, Antonio Pineda and 
Yukinari Sumino for numerous informative discussions. This work has been 
supported by the DFG and the NSFC through funds provided to the Sino-German CRC 
110 ``Symmetries and the Emergence of Structure in QCD'', and by the DFG 
cluster of excellence ``Origin and structure of the universe'' 
(\url{www.universe-cluster.de}). J.S. acknowledges the financial support from 
the Alexander von Humboldt Foundation and thanks the Technische Universit\"at 
M\"unchen for hospitality while most of this work was carried out.
\end{acknowledgments}

\appendix

\section{Solving the Schr\"odinger equation}
\label{app:SolvingSchroedingerEquation}
In a generic central potential, $V(r)$, the Schr\"odinger equation for the 
reduced wave function, $u_{n\ell}(r) = r R_{n\ell}(r)$, has the form
\begin{equation}
\label{app:eq:1dim}
-\frac{1}{m} \frac{\d^2 u_{n\ell}(r)}{\d r^2} + \left( V(r) + 
\frac{\ell(\ell+1)}{m r^2} \right) u_{n\ell}(r) = E_{nl} u_{n\ell}(r) \,,
\end{equation}
in the case of two particles of mass $m$. This is a one dimensional 
Schr\"odinger equation, which has significance only for positive values of $r$,
and must be supplemented by a boundary condition at $r=0$. We require that the 
radial function $R_{n\ell}(r)$ remains finite at the origin, which implies that 
$u_{n\ell}(0) = 0$.

If close to the origin the potential $V(r)$ has the form
\begin{equation}
\label{app:eq:PolyPot}
V(r) = r^p (b_0 + b_1 r + \ldots) \,, \quad b_{0} \neq 0 \,,
\end{equation}
where $p$ is an integer such that $p\geq-1$, we can expand the solution 
$u_{n\ell}(r)$ in the vicinity of the origin as
\begin{equation}
u_{n\ell}(r) = r^s \sum_{k=0}^{\infty} c_k r^k \,, \quad c_{0} \neq 0 \,.
\end{equation}
Equation~\eqref{app:eq:1dim} requires that $s(s-1)-\ell(\ell+1)=0$, so that 
$s=\ell+1$ or $s=-\ell$. The choice $s=-\ell$ corresponds to irregular 
solutions that do not satisfy the boundary condition $u_{n\ell}(0) = 0$. The 
other choice $s=\ell+1$ corresponds to regular solutions that are physically 
allowed, and are such that
\begin{equation}
u_{n\ell}(r) \stackrel{r\to 0}{\sim} r^{\ell+1} \,.
\end{equation}
Since we are interested in finding bound states, we also impose that 
\begin{equation}
\label{app:eq:aymptotic}
u_{n\ell}(r) \stackrel{r\to \infty}{\sim} e^{-kr} \,, 
\end{equation}
where $k = \sqrt{m|E_{n\ell}|}$ is the wave function number.

\begin{figure}[ht]
\centering
\includegraphics[width=0.42\textwidth]{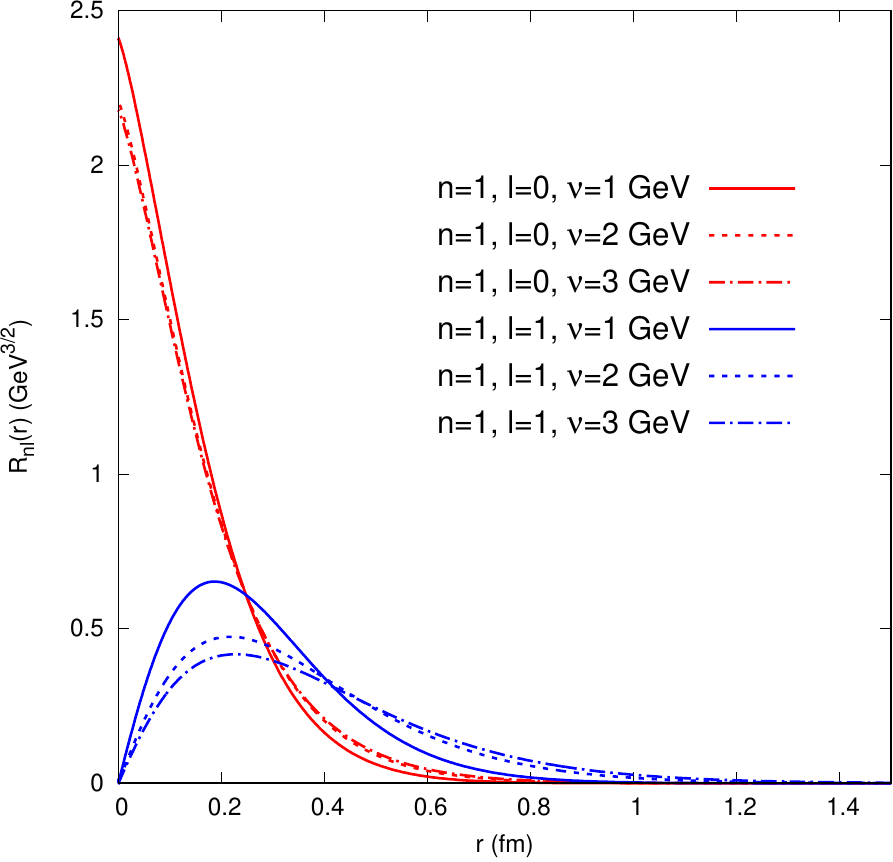}
\caption{\label{fig:wf} Radial wave functions, $R_{n\ell}(r)$, for different 
values of $n$, $\ell$, and $\nu$ as a function of $r$. $R_{n\ell}(r)$ is the 
(numerical) solution of the Schr\"odinger equation for the 
potential~\eqref{eq:RSpScheme}.}
\end{figure}

\begin{figure}[h]
\centering
\includegraphics[width=0.4\textwidth]{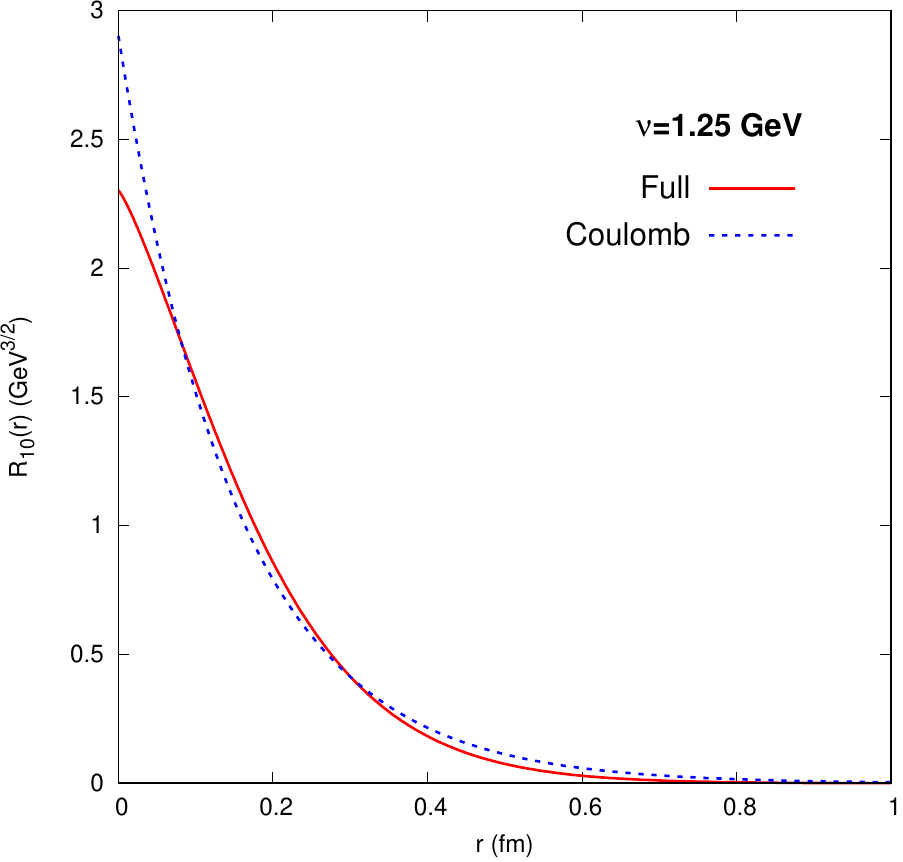}
\caption{\label{fig:wfS} Radial wave function, $R_{10}(r)$, of the lowest 
$S$-wave state for $\nu=1.25$~GeV as a function of $r$. The red solid line 
shows the (numerical) solution of the Schr\"odinger equation for the 
potential~\eqref{eq:RSpScheme}, while the blue dotted line shows the  solution 
of the Schr\"odinger equation for the leading order Coulomb 
potential~\eqref{LOCoulomb}.}
\end{figure}

In numerical applications we introduce short- and long-distance cut-offs, 
denoted by $r_{\text{in}}$ and $r_{\text{fi}}$, respectively, for which we 
require 
\begin{align}
u_{n\ell}(r_{\text{in}}) = r_{\text{in}}^{\ell+1} \,, \label{cond1}\\
u_{n\ell}(r_{\text{fi}}) = e^{-kr_\text{fi}} \label{cond2}\,.
\end{align}
The dependence of physical observables on the short distance cut-off is quite 
sensible, and it can hinder the numerical search of the ground state and its 
excitations due to the dominance of the irregular solutions at very small 
values of $r$. In order to improve on this, we can use, for two different 
energies $E_{n\ell} \neq E_{n'\ell}$, the orthogonality relation between their 
bound state wave functions:
\begin{align}
\label{app:eq:orthogonal}
u_{n\ell}'(r_{\text{in}}) u_{n'\ell}(r_{\text{in}}) &- u_{n\ell}(r_{\text{in}}) 
u_{n'\ell}'(r_{\text{in}}) \nonumber\\
& \hspace{-1cm}= m (E_{n\ell}-E_{n'\ell}) \int_{r_{\text{in}}}^{\infty} \d r \, 
u_{n\ell}(r) u_{n'\ell}(r) \,,
\end{align}
which follows from multiplying Eq.~\eqref{app:eq:1dim} by $u_{n'\ell}(r)$ and 
later subtracting the same equation, but with $n$ and $n'$ exchanged. The 
regularity condition at the origin $u_{n\ell}(r_{\text{in}})=0$ for 
$r_{\text{in}}\to 0$ makes the states automatically orthogonal in the 
$r_{\text{in}}\to 0$ limit. We can further enforce orthogonality also for 
finite $r_{\text{in}}$ by requiring
\begin{equation}
\label{app:eq:matching}
\frac{u_{n\ell}'(r_{\text{in}})}{u_{n\ell}(r_{\text{in}})} = 
\frac{u_{n'\ell}'(r_{\text{in}})}{u_{n'\ell}(r_{\text{in}})} \,,
\end{equation}
for any two states, meaning that the logarithmic derivative at short distances 
becomes independent of the principal quantum number. This condition has many 
advantages, such as the possibility of working with singular potentials at 
$r=0$, like $r^2 V(r) = \pm \infty$ for $r\to 0$. Moreover, in order to avoid 
pollution from the irregular solutions, we can use Eq.~\eqref{app:eq:matching} 
to match at an intermediate distance $r_\text{me}$ the solutions of the one 
dimensional Schr\"odinger equation obtained when integrating it from 
$r_\text{in}$ to $r_\text{me}$, with boundary \eqref{cond1}, and from 
$r_\text{me}$ to $r_\text{fi}$ with boundary \eqref{cond2}. Finally, 
Eq.~\eqref{app:eq:matching} is ideal to find excited states because, as we 
remarked, the logarithmic derivative at short distances becomes independent of 
$n$. For further details we refer to Ref.~\cite{Segovia:2011tb}.

\begin{figure}[ht]
\centering
\includegraphics[width=0.4\textwidth]{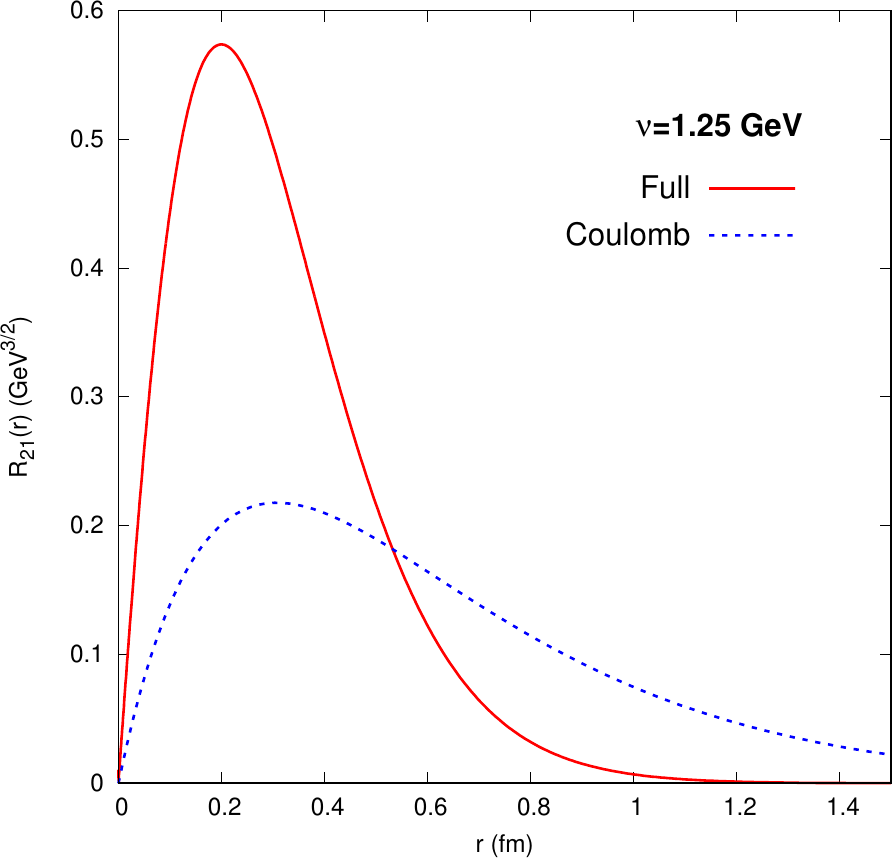}
\caption{\label{fig:wfP} As in Fig.~\ref{fig:wfS}, but for the radial wave 
function, $R_{21}(r)$, of the lowest $P$-wave state.}
\end{figure}

In order to solve the differential equation~\eqref{app:eq:1dim} for the 
potential~\eqref{eq:RSpScheme}, we use the fourth order Runge--Kutta algorithm 
with adaptive step size implemented in FORTRAN77. This implementation 
automatically takes care of convergence and numerical accuracy. The numerical 
implementation of the Green function, Eqs.~\eqref{eq:GreenFunctionI} 
and~\eqref{eq:GreenFunctionII}, involves a sum over intermediate states. We 
compute as many intermediate states, and include them, until we see 
convergence. Solutions for the radial wave functions, $R_{n\ell}(r)$, are shown 
in the Figs.~\ref{fig:wf}, \ref{fig:wfS}, and~\ref{fig:wfP}, where we also 
compare with the leading order Coulomb wave functions.

\bibliography{E1TransitionsInpNRQCD}

\begin{thebibliography}{50}%
\makeatletter
\providecommand \@ifxundefined [1]{%
 \@ifx{#1\undefined}
}%
\providecommand \@ifnum [1]{%
 \ifnum #1\expandafter \@firstoftwo
 \else \expandafter \@secondoftwo
 \fi
}%
\providecommand \@ifx [1]{%
 \ifx #1\expandafter \@firstoftwo
 \else \expandafter \@secondoftwo
 \fi
}%
\providecommand \natexlab [1]{#1}%
\providecommand \enquote  [1]{``#1''}%
\providecommand \bibnamefont  [1]{#1}%
\providecommand \bibfnamefont [1]{#1}%
\providecommand \citenamefont [1]{#1}%
\providecommand \href@noop [0]{\@secondoftwo}%
\providecommand \href [0]{\begingroup \@sanitize@url \@href}%
\providecommand \@href[1]{\@@startlink{#1}\@@href}%
\providecommand \@@href[1]{\endgroup#1\@@endlink}%
\providecommand \@sanitize@url [0]{\catcode `\\12\catcode `\$12\catcode
  `\&12\catcode `\#12\catcode `\^12\catcode `\_12\catcode `\%12\relax}%
\providecommand \@@startlink[1]{}%
\providecommand \@@endlink[0]{}%
\providecommand \url  [0]{\begingroup\@sanitize@url \@url }%
\providecommand \@url [1]{\endgroup\@href {#1}{\urlprefix }}%
\providecommand \urlprefix  [0]{URL }%
\providecommand \Eprint [0]{\href }%
\@ifxundefined \urlstyle {%
  \providecommand \doi  [0]{\begingroup \@sanitize@url \@doi}%
  \providecommand \@doi [1]{\endgroup \@@startlink {\doibase
  #1}doi:\discretionary {}{}{}#1\@@endlink }%
}{%
  \providecommand \doi  [0]{doi:\discretionary{}{}{}\begingroup
  \urlstyle{rm}\Url }%
}%
\providecommand \doibase [0]{http://dx.doi.org/}%
\providecommand \Doi [0]{\begingroup \@sanitize@url \@Doi }%
\providecommand \@Doi  [1]{\endgroup\@@startlink{\doibase#1}\@@Doi}%
\providecommand \@@Doi [1]{#1\@@endlink}%
\providecommand \selectlanguage [0]{\@gobble}%
\providecommand \bibinfo  [0]{\@secondoftwo}%
\providecommand \bibfield  [0]{\@secondoftwo}%
\providecommand \translation [1]{[#1]}%
\providecommand \BibitemOpen [0]{}%
\providecommand \bibitemStop [0]{}%
\providecommand \bibitemNoStop [0]{.\EOS\space}%
\providecommand \EOS [0]{\spacefactor3000\relax}%
\providecommand \BibitemShut  [1]{\csname bibitem#1\endcsname}%
\bibitem [{\citenamefont {Han}\ \emph {et~al.}(1982)\citenamefont {Han} \emph
  {et~al.}}]{Han:1982zk}%
  \BibitemOpen
  \bibfield  {author} {\bibinfo {author} {\bibfnamefont {K.}~\bibnamefont
  {Han}} \emph {et~al.},\ }\Doi {10.1103/PhysRevLett.49.1612} {\bibfield
  {journal} {\bibinfo  {journal} {Phys. Rev. Lett.},\ }\textbf {\bibinfo
  {volume} {49}},\ \bibinfo {pages} {1612} (\bibinfo {year}
  {1982})}\BibitemShut {NoStop}%
\bibitem [{\citenamefont {Eigen}\ \emph {et~al.}(1982)\citenamefont {Eigen}
  \emph {et~al.}}]{Eigen:1982zm}%
  \BibitemOpen
  \bibfield  {author} {\bibinfo {author} {\bibfnamefont {G.}~\bibnamefont
  {Eigen}} \emph {et~al.},\ }\Doi {10.1103/PhysRevLett.49.1616} {\bibfield
  {journal} {\bibinfo  {journal} {Phys. Rev. Lett.},\ }\textbf {\bibinfo
  {volume} {49}},\ \bibinfo {pages} {1616} (\bibinfo {year}
  {1982})}\BibitemShut {NoStop}%
\bibitem [{\citenamefont {Klopfenstein}\ \emph {et~al.}(1983)\citenamefont
  {Klopfenstein} \emph {et~al.}}]{Klopfenstein:1983nx}%
  \BibitemOpen
  \bibfield  {author} {\bibinfo {author} {\bibfnamefont {C.}~\bibnamefont
  {Klopfenstein}} \emph {et~al.},\ }\bibfield  {booktitle} {\emph {\bibinfo
  {booktitle} {{11th International Symposium on Lepton and Photon Interactions
  at High Energies Ithaca, New York, August 4-9, 1983}}},\ }\Doi
  {10.1103/PhysRevLett.51.160} {\bibfield  {journal} {\bibinfo  {journal}
  {Phys. Rev. Lett.},\ }\textbf {\bibinfo {volume} {51}},\ \bibinfo {pages}
  {160} (\bibinfo {year} {1983})}\BibitemShut {NoStop}%
\bibitem [{\citenamefont {Pauss}\ \emph {et~al.}(1983)\citenamefont {Pauss}
  \emph {et~al.}}]{Pauss:1983pa}%
  \BibitemOpen
  \bibfield  {author} {\bibinfo {author} {\bibfnamefont {F.}~\bibnamefont
  {Pauss}} \emph {et~al.},\ }\bibfield  {booktitle} {\emph {\bibinfo
  {booktitle} {{11th International Symposium on Lepton and Photon Interactions
  at High Energies Ithaca, New York, August 4-9, 1983}}},\ }\Doi
  {10.1016/0370-2693(83)91539-3} {\bibfield  {journal} {\bibinfo  {journal}
  {Phys. Lett.},\ }\textbf {\bibinfo {volume} {B130}},\ \bibinfo {pages} {439}
  (\bibinfo {year} {1983})}\BibitemShut {NoStop}%
\bibitem [{\citenamefont {Brambilla}\ \emph {et~al.}(2004)\citenamefont
  {Brambilla} \emph {et~al.}}]{Brambilla:2004wf}%
  \BibitemOpen
  \bibfield  {author} {\bibinfo {author} {\bibfnamefont {N.}~\bibnamefont
  {Brambilla}} \emph {et~al.} (\bibinfo {collaboration} {Quarkonium Working
  Group}),\ }\href@noop {} { (\bibinfo {year} {2004})},\ \Eprint
  {http://arxiv.org/abs/hep-ph/0412158} {arXiv:hep-ph/0412158 [hep-ph]}
  \BibitemShut {NoStop}%
\bibitem [{\citenamefont {Tanabashi}\ \emph {et~al.}(2018)\citenamefont
  {Tanabashi} \emph {et~al.}}]{PhysRevD.98.030001}%
  \BibitemOpen
  \bibfield  {author} {\bibinfo {author} {\bibfnamefont {M.}~\bibnamefont
  {Tanabashi}} \emph {et~al.} (\bibinfo {collaboration} {Particle Data
  Group}),\ }\Doi {10.1103/PhysRevD.98.030001} {\bibfield  {journal} {\bibinfo
  {journal} {Phys. Rev. D},\ }\textbf {\bibinfo {volume} {98}},\ \bibinfo
  {pages} {030001} (\bibinfo {year} {2018})}\BibitemShut {NoStop}%
\bibitem [{\citenamefont {Segovia}\ \emph {et~al.}(2016)\citenamefont
  {Segovia}, \citenamefont {Ortega}, \citenamefont {Entem},\ and\ \citenamefont
  {Fern{\'a}ndez}}]{Segovia:2016xqb}%
  \BibitemOpen
  \bibfield  {author} {\bibinfo {author} {\bibfnamefont {J.}~\bibnamefont
  {Segovia}}, \bibinfo {author} {\bibfnamefont {P.~G.}\ \bibnamefont {Ortega}},
  \bibinfo {author} {\bibfnamefont {D.~R.}\ \bibnamefont {Entem}}, \ and\
  \bibinfo {author} {\bibfnamefont {F.}~\bibnamefont {Fern{\'a}ndez}},\ }\Doi
  {10.1103/PhysRevD.93.074027} {\bibfield  {journal} {\bibinfo  {journal}
  {Phys. Rev.},\ }\textbf {\bibinfo {volume} {D93}},\ \bibinfo {pages} {074027}
  (\bibinfo {year} {2016})},\ \Eprint {http://arxiv.org/abs/1601.05093}
  {arXiv:1601.05093 [hep-ph]} \BibitemShut {NoStop}%
\bibitem [{\citenamefont {Brambilla}\ \emph {et~al.}(2011)\citenamefont
  {Brambilla} \emph {et~al.}}]{Brambilla:2010cs}%
  \BibitemOpen
  \bibfield  {author} {\bibinfo {author} {\bibfnamefont {N.}~\bibnamefont
  {Brambilla}} \emph {et~al.},\ }\Doi {10.1140/epjc/s10052-010-1534-9}
  {\bibfield  {journal} {\bibinfo  {journal} {Eur. Phys. J.},\ }\textbf
  {\bibinfo {volume} {C71}},\ \bibinfo {pages} {1534} (\bibinfo {year}
  {2011})},\ \Eprint {http://arxiv.org/abs/1010.5827} {arXiv:1010.5827
  [hep-ph]} \BibitemShut {NoStop}%
\bibitem [{\citenamefont {Brambilla}\ \emph {et~al.}(2014)\citenamefont
  {Brambilla} \emph {et~al.}}]{Brambilla:2014jmp}%
  \BibitemOpen
  \bibfield  {author} {\bibinfo {author} {\bibfnamefont {N.}~\bibnamefont
  {Brambilla}} \emph {et~al.},\ }\Doi {10.1140/epjc/s10052-014-2981-5}
  {\bibfield  {journal} {\bibinfo  {journal} {Eur. Phys. J.},\ }\textbf
  {\bibinfo {volume} {C74}},\ \bibinfo {pages} {2981} (\bibinfo {year}
  {2014})},\ \Eprint {http://arxiv.org/abs/1404.3723} {arXiv:1404.3723
  [hep-ph]} \BibitemShut {NoStop}%
\bibitem [{\citenamefont {Pineda}\ and\ \citenamefont
  {Soto}(1998)}]{Pineda:1997bj}%
  \BibitemOpen
  \bibfield  {author} {\bibinfo {author} {\bibfnamefont {A.}~\bibnamefont
  {Pineda}}\ and\ \bibinfo {author} {\bibfnamefont {J.}~\bibnamefont {Soto}},\
  }\bibfield  {booktitle} {\emph {\bibinfo {booktitle} {{Quantum
  chromodynamics. Proceedings, Conference, QCD'97, Montpellier, France, July
  3-9, 1997}}},\ }\Doi {10.1016/S0920-5632(97)01102-X} {\bibfield  {journal}
  {\bibinfo  {journal} {Nucl. Phys. Proc. Suppl.},\ }\textbf {\bibinfo {volume}
  {64}},\ \bibinfo {pages} {428} (\bibinfo {year} {1998})},\ \bibinfo {note}
  {[,428(1997)]},\ \Eprint {http://arxiv.org/abs/hep-ph/9707481}
  {arXiv:hep-ph/9707481 [hep-ph]} \BibitemShut {NoStop}%
\bibitem [{\citenamefont {Brambilla}\ \emph {et~al.}(2000)\citenamefont
  {Brambilla}, \citenamefont {Pineda}, \citenamefont {Soto},\ and\
  \citenamefont {Vairo}}]{Brambilla:1999xf}%
  \BibitemOpen
  \bibfield  {author} {\bibinfo {author} {\bibfnamefont {N.}~\bibnamefont
  {Brambilla}}, \bibinfo {author} {\bibfnamefont {A.}~\bibnamefont {Pineda}},
  \bibinfo {author} {\bibfnamefont {J.}~\bibnamefont {Soto}}, \ and\ \bibinfo
  {author} {\bibfnamefont {A.}~\bibnamefont {Vairo}},\ }\Doi
  {10.1016/S0550-3213(99)00693-8} {\bibfield  {journal} {\bibinfo  {journal}
  {Nucl. Phys.},\ }\textbf {\bibinfo {volume} {B566}},\ \bibinfo {pages} {275}
  (\bibinfo {year} {2000})},\ \Eprint {http://arxiv.org/abs/hep-ph/9907240}
  {arXiv:hep-ph/9907240 [hep-ph]} \BibitemShut {NoStop}%
\bibitem [{\citenamefont {Brambilla}\ \emph {et~al.}(2005)\citenamefont
  {Brambilla}, \citenamefont {Pineda}, \citenamefont {Soto},\ and\
  \citenamefont {Vairo}}]{Brambilla:2004jw}%
  \BibitemOpen
  \bibfield  {author} {\bibinfo {author} {\bibfnamefont {N.}~\bibnamefont
  {Brambilla}}, \bibinfo {author} {\bibfnamefont {A.}~\bibnamefont {Pineda}},
  \bibinfo {author} {\bibfnamefont {J.}~\bibnamefont {Soto}}, \ and\ \bibinfo
  {author} {\bibfnamefont {A.}~\bibnamefont {Vairo}},\ }\Doi
  {10.1103/RevModPhys.77.1423} {\bibfield  {journal} {\bibinfo  {journal} {Rev.
  Mod. Phys.},\ }\textbf {\bibinfo {volume} {77}},\ \bibinfo {pages} {1423}
  (\bibinfo {year} {2005})},\ \Eprint {http://arxiv.org/abs/hep-ph/0410047}
  {arXiv:hep-ph/0410047 [hep-ph]} \BibitemShut {NoStop}%
\bibitem [{\citenamefont {Pineda}(2012)}]{Pineda:2011dg}%
  \BibitemOpen
  \bibfield  {author} {\bibinfo {author} {\bibfnamefont {A.}~\bibnamefont
  {Pineda}},\ }\Doi {10.1016/j.ppnp.2012.01.038} {\bibfield  {journal}
  {\bibinfo  {journal} {Prog. Part. Nucl. Phys.},\ }\textbf {\bibinfo {volume}
  {67}},\ \bibinfo {pages} {735} (\bibinfo {year} {2012})}\BibitemShut
  {NoStop}%
\bibitem [{\citenamefont {Brambilla}\ \emph {et~al.}(2006)\citenamefont
  {Brambilla}, \citenamefont {Jia},\ and\ \citenamefont
  {Vairo}}]{Brambilla:2005zw}%
  \BibitemOpen
  \bibfield  {author} {\bibinfo {author} {\bibfnamefont {N.}~\bibnamefont
  {Brambilla}}, \bibinfo {author} {\bibfnamefont {Y.}~\bibnamefont {Jia}}, \
  and\ \bibinfo {author} {\bibfnamefont {A.}~\bibnamefont {Vairo}},\ }\Doi
  {10.1103/PhysRevD.73.054005} {\bibfield  {journal} {\bibinfo  {journal}
  {Phys. Rev.},\ }\textbf {\bibinfo {volume} {D73}},\ \bibinfo {pages} {054005}
  (\bibinfo {year} {2006})},\ \Eprint {http://arxiv.org/abs/hep-ph/0512369}
  {arXiv:hep-ph/0512369 [hep-ph]} \BibitemShut {NoStop}%
\bibitem [{\citenamefont {Brambilla}\ \emph {et~al.}(2012)\citenamefont
  {Brambilla}, \citenamefont {Pietrulewicz},\ and\ \citenamefont
  {Vairo}}]{Brambilla:2012be}%
  \BibitemOpen
  \bibfield  {author} {\bibinfo {author} {\bibfnamefont {N.}~\bibnamefont
  {Brambilla}}, \bibinfo {author} {\bibfnamefont {P.}~\bibnamefont
  {Pietrulewicz}}, \ and\ \bibinfo {author} {\bibfnamefont {A.}~\bibnamefont
  {Vairo}},\ }\Doi {10.1103/PhysRevD.85.094005} {\bibfield  {journal} {\bibinfo
   {journal} {Phys. Rev.},\ }\textbf {\bibinfo {volume} {D85}},\ \bibinfo
  {pages} {094005} (\bibinfo {year} {2012})},\ \Eprint
  {http://arxiv.org/abs/1203.3020} {arXiv:1203.3020 [hep-ph]} \BibitemShut
  {NoStop}%
\bibitem [{\citenamefont {Brambilla}\ \emph {et~al.}(2001)\citenamefont
  {Brambilla}, \citenamefont {Sumino},\ and\ \citenamefont
  {Vairo}}]{Brambilla:2001fw}%
  \BibitemOpen
  \bibfield  {author} {\bibinfo {author} {\bibfnamefont {N.}~\bibnamefont
  {Brambilla}}, \bibinfo {author} {\bibfnamefont {Y.}~\bibnamefont {Sumino}}, \
  and\ \bibinfo {author} {\bibfnamefont {A.}~\bibnamefont {Vairo}},\ }\Doi
  {10.1016/S0370-2693(01)00611-6} {\bibfield  {journal} {\bibinfo  {journal}
  {Phys. Lett.},\ }\textbf {\bibinfo {volume} {B513}},\ \bibinfo {pages} {381}
  (\bibinfo {year} {2001})},\ \Eprint {http://arxiv.org/abs/hep-ph/0101305}
  {arXiv:hep-ph/0101305 [hep-ph]} \BibitemShut {NoStop}%
\bibitem [{\citenamefont {Brambilla}\ \emph {et~al.}(2002)\citenamefont
  {Brambilla}, \citenamefont {Sumino},\ and\ \citenamefont
  {Vairo}}]{Brambilla:2001qk}%
  \BibitemOpen
  \bibfield  {author} {\bibinfo {author} {\bibfnamefont {N.}~\bibnamefont
  {Brambilla}}, \bibinfo {author} {\bibfnamefont {Y.}~\bibnamefont {Sumino}}, \
  and\ \bibinfo {author} {\bibfnamefont {A.}~\bibnamefont {Vairo}},\ }\Doi
  {10.1103/PhysRevD.65.034001} {\bibfield  {journal} {\bibinfo  {journal}
  {Phys. Rev.},\ }\textbf {\bibinfo {volume} {D65}},\ \bibinfo {pages} {034001}
  (\bibinfo {year} {2002})},\ \Eprint {http://arxiv.org/abs/hep-ph/0108084}
  {arXiv:hep-ph/0108084 [hep-ph]} \BibitemShut {NoStop}%
\bibitem [{\citenamefont {Brambilla}\ and\ \citenamefont
  {Vairo}(2005)}]{Brambilla:2004wu}%
  \BibitemOpen
  \bibfield  {author} {\bibinfo {author} {\bibfnamefont {N.}~\bibnamefont
  {Brambilla}}\ and\ \bibinfo {author} {\bibfnamefont {A.}~\bibnamefont
  {Vairo}},\ }\Doi {10.1103/PhysRevD.71.034020} {\bibfield  {journal} {\bibinfo
   {journal} {Phys. Rev.},\ }\textbf {\bibinfo {volume} {D71}},\ \bibinfo
  {pages} {034020} (\bibinfo {year} {2005})},\ \Eprint
  {http://arxiv.org/abs/hep-ph/0411156} {arXiv:hep-ph/0411156 [hep-ph]}
  \BibitemShut {NoStop}%
\bibitem [{\citenamefont {Sumino}(2016)}]{Sumino:2016sxe}%
  \BibitemOpen
  \bibfield  {author} {\bibinfo {author} {\bibfnamefont {Y.}~\bibnamefont
  {Sumino}},\ }\bibfield  {booktitle} {\emph {\bibinfo {booktitle}
  {{Proceedings, 13th DESY Workshop on Elementary Particle Physics: Loops and
  Legs in Quantum Field Theory (LL2016): Leipzig, Germany, April 24-29,
  2016}}},\ }\Doi {10.22323/1.260.0011} {\bibfield  {journal} {\bibinfo
  {journal} {PoS},\ }\textbf {\bibinfo {volume} {LL2016}},\ \bibinfo {pages}
  {011} (\bibinfo {year} {2016})},\ \Eprint {http://arxiv.org/abs/1607.03469}
  {arXiv:1607.03469 [hep-ph]} \BibitemShut {NoStop}%
\bibitem [{\citenamefont {Peset}\ \emph {et~al.}(2018)\citenamefont {Peset},
  \citenamefont {Pineda},\ and\ \citenamefont {Segovia}}]{Peset:2018jkf}%
  \BibitemOpen
  \bibfield  {author} {\bibinfo {author} {\bibfnamefont {C.}~\bibnamefont
  {Peset}}, \bibinfo {author} {\bibfnamefont {A.}~\bibnamefont {Pineda}}, \
  and\ \bibinfo {author} {\bibfnamefont {J.}~\bibnamefont {Segovia}},\ }\Doi
  {10.1103/PhysRevD.98.094003} {\bibfield  {journal} {\bibinfo  {journal}
  {Phys. Rev.},\ }\textbf {\bibinfo {volume} {D98}},\ \bibinfo {pages} {094003}
  (\bibinfo {year} {2018})},\ \Eprint {http://arxiv.org/abs/1809.09124}
  {arXiv:1809.09124 [hep-ph]} \BibitemShut {NoStop}%
\bibitem [{\citenamefont {Pineda}(2001)}]{Pineda:2001zq}%
  \BibitemOpen
  \bibfield  {author} {\bibinfo {author} {\bibfnamefont {A.}~\bibnamefont
  {Pineda}},\ }\Doi {10.1088/1126-6708/2001/06/022} {\bibfield  {journal}
  {\bibinfo  {journal} {JHEP},\ }\textbf {\bibinfo {volume} {06}},\ \bibinfo
  {pages} {022} (\bibinfo {year} {2001})},\ \Eprint
  {http://arxiv.org/abs/hep-ph/0105008} {arXiv:hep-ph/0105008 [hep-ph]}
  \BibitemShut {NoStop}%
\bibitem [{\citenamefont {Kiyo}\ \emph {et~al.}(2010)\citenamefont {Kiyo},
  \citenamefont {Pineda},\ and\ \citenamefont {Signer}}]{Kiyo:2010jm}%
  \BibitemOpen
  \bibfield  {author} {\bibinfo {author} {\bibfnamefont {Y.}~\bibnamefont
  {Kiyo}}, \bibinfo {author} {\bibfnamefont {A.}~\bibnamefont {Pineda}}, \ and\
  \bibinfo {author} {\bibfnamefont {A.}~\bibnamefont {Signer}},\ }\Doi
  {10.1016/j.nuclphysb.2010.08.007} {\bibfield  {journal} {\bibinfo  {journal}
  {Nucl. Phys.},\ }\textbf {\bibinfo {volume} {B841}},\ \bibinfo {pages} {231}
  (\bibinfo {year} {2010})},\ \Eprint {http://arxiv.org/abs/1006.2685}
  {arXiv:1006.2685 [hep-ph]} \BibitemShut {NoStop}%
\bibitem [{\citenamefont {Pineda}\ and\ \citenamefont
  {Segovia}(2013)}]{Pineda:2013lta}%
  \BibitemOpen
  \bibfield  {author} {\bibinfo {author} {\bibfnamefont {A.}~\bibnamefont
  {Pineda}}\ and\ \bibinfo {author} {\bibfnamefont {J.}~\bibnamefont
  {Segovia}},\ }\Doi {10.1103/PhysRevD.87.074024} {\bibfield  {journal}
  {\bibinfo  {journal} {Phys. Rev.},\ }\textbf {\bibinfo {volume} {D87}},\
  \bibinfo {pages} {074024} (\bibinfo {year} {2013})},\ \Eprint
  {http://arxiv.org/abs/1302.3528} {arXiv:1302.3528 [hep-ph]} \BibitemShut
  {NoStop}%
\bibitem [{\citenamefont {Pietrulewicz}(2013)}]{Pietrulewicz:2013ct}%
  \BibitemOpen
  \bibfield  {author} {\bibinfo {author} {\bibfnamefont {P.}~\bibnamefont
  {Pietrulewicz}},\ }\bibfield  {booktitle} {\emph {\bibinfo {booktitle}
  {{Proceedings, 10th Conference on Quark Confinement and the Hadron Spectrum
  (Confinement X): Munich, Germany, October 8-12, 2012}}},\ }\Doi
  {10.22323/1.171.0135} { (\bibinfo {year} {2013})},\ \doi
  {10.22323/1.171.0135},\ \bibinfo {note} {[PoSConfinementX,135(2012)]},\
  \Eprint {http://arxiv.org/abs/1301.1308} {arXiv:1301.1308 [hep-ph]}
  \BibitemShut {NoStop}%
\bibitem [{\citenamefont {Martinez}(2016)}]{Martinez:2016spe}%
  \BibitemOpen
  \bibfield  {author} {\bibinfo {author} {\bibfnamefont {H.~E.}\ \bibnamefont
  {Martinez}},\ }\bibfield  {booktitle} {\emph {\bibinfo {booktitle}
  {{Proceedings, 11th Conference on Quark Confinement and the Hadron Spectrum
  (Confinement XI): St. Petersburg, Russia, September 8-12, 2014}}},\ }\Doi
  {10.1063/1.4938648} {\bibfield  {journal} {\bibinfo  {journal} {AIP Conf.
  Proc.},\ }\textbf {\bibinfo {volume} {1701}},\ \bibinfo {pages} {050008}
  (\bibinfo {year} {2016})}\BibitemShut {NoStop}%
\bibitem [{\citenamefont {Fischler}(1977)}]{Fischler:1977yf}%
  \BibitemOpen
  \bibfield  {author} {\bibinfo {author} {\bibfnamefont {W.}~\bibnamefont
  {Fischler}},\ }\Doi {10.1016/0550-3213(77)90026-8} {\bibfield  {journal}
  {\bibinfo  {journal} {Nucl. Phys.},\ }\textbf {\bibinfo {volume} {B129}},\
  \bibinfo {pages} {157} (\bibinfo {year} {1977})}\BibitemShut {NoStop}%
\bibitem [{\citenamefont {Schr{\"o}der}(1999)}]{Schroder:1998vy}%
  \BibitemOpen
  \bibfield  {author} {\bibinfo {author} {\bibfnamefont {Y.}~\bibnamefont
  {Schr{\"o}der}},\ }\Doi {10.1016/S0370-2693(99)00010-6} {\bibfield  {journal}
  {\bibinfo  {journal} {Phys. Lett.},\ }\textbf {\bibinfo {volume} {B447}},\
  \bibinfo {pages} {321} (\bibinfo {year} {1999})},\ \Eprint
  {http://arxiv.org/abs/hep-ph/9812205} {arXiv:hep-ph/9812205 [hep-ph]}
  \BibitemShut {NoStop}%
\bibitem [{\citenamefont {Brambilla}\ \emph {et~al.}(1999)\citenamefont
  {Brambilla}, \citenamefont {Pineda}, \citenamefont {Soto},\ and\
  \citenamefont {Vairo}}]{Brambilla:1999qa}%
  \BibitemOpen
  \bibfield  {author} {\bibinfo {author} {\bibfnamefont {N.}~\bibnamefont
  {Brambilla}}, \bibinfo {author} {\bibfnamefont {A.}~\bibnamefont {Pineda}},
  \bibinfo {author} {\bibfnamefont {J.}~\bibnamefont {Soto}}, \ and\ \bibinfo
  {author} {\bibfnamefont {A.}~\bibnamefont {Vairo}},\ }\Doi
  {10.1103/PhysRevD.60.091502} {\bibfield  {journal} {\bibinfo  {journal}
  {Phys. Rev.},\ }\textbf {\bibinfo {volume} {D60}},\ \bibinfo {pages} {091502}
  (\bibinfo {year} {1999})},\ \Eprint {http://arxiv.org/abs/hep-ph/9903355}
  {arXiv:hep-ph/9903355 [hep-ph]} \BibitemShut {NoStop}%
\bibitem [{\citenamefont {Anzai}\ \emph {et~al.}(2010)\citenamefont {Anzai},
  \citenamefont {Kiyo},\ and\ \citenamefont {Sumino}}]{Anzai:2009tm}%
  \BibitemOpen
  \bibfield  {author} {\bibinfo {author} {\bibfnamefont {C.}~\bibnamefont
  {Anzai}}, \bibinfo {author} {\bibfnamefont {Y.}~\bibnamefont {Kiyo}}, \ and\
  \bibinfo {author} {\bibfnamefont {Y.}~\bibnamefont {Sumino}},\ }\Doi
  {10.1103/PhysRevLett.104.112003} {\bibfield  {journal} {\bibinfo  {journal}
  {Phys. Rev. Lett.},\ }\textbf {\bibinfo {volume} {104}},\ \bibinfo {pages}
  {112003} (\bibinfo {year} {2010})},\ \Eprint {http://arxiv.org/abs/0911.4335}
  {arXiv:0911.4335 [hep-ph]} \BibitemShut {NoStop}%
\bibitem [{\citenamefont {Smirnov}\ \emph {et~al.}(2010)\citenamefont
  {Smirnov}, \citenamefont {Smirnov},\ and\ \citenamefont
  {Steinhauser}}]{Smirnov:2009fh}%
  \BibitemOpen
  \bibfield  {author} {\bibinfo {author} {\bibfnamefont {A.~V.}\ \bibnamefont
  {Smirnov}}, \bibinfo {author} {\bibfnamefont {V.~A.}\ \bibnamefont
  {Smirnov}}, \ and\ \bibinfo {author} {\bibfnamefont {M.}~\bibnamefont
  {Steinhauser}},\ }\Doi {10.1103/PhysRevLett.104.112002} {\bibfield  {journal}
  {\bibinfo  {journal} {Phys. Rev. Lett.},\ }\textbf {\bibinfo {volume}
  {104}},\ \bibinfo {pages} {112002} (\bibinfo {year} {2010})},\ \Eprint
  {http://arxiv.org/abs/0911.4742} {arXiv:0911.4742 [hep-ph]} \BibitemShut
  {NoStop}%
\bibitem [{\citenamefont {Voloshin}(1982)}]{Voloshin:1979uv}%
  \BibitemOpen
  \bibfield  {author} {\bibinfo {author} {\bibfnamefont {M.~B.}\ \bibnamefont
  {Voloshin}},\ }\href@noop {} {\bibfield  {journal} {\bibinfo  {journal} {Sov.
  J. Nucl. Phys.},\ }\textbf {\bibinfo {volume} {36}},\ \bibinfo {pages} {143}
  (\bibinfo {year} {1982})},\ \bibinfo {note} {[Yad.
  Fiz.36,247(1982)]}\BibitemShut {NoStop}%
\bibitem [{\citenamefont {Kiyo}\ and\ \citenamefont
  {Sumino}(2014)}]{Kiyo:2014uca}%
  \BibitemOpen
  \bibfield  {author} {\bibinfo {author} {\bibfnamefont {Y.}~\bibnamefont
  {Kiyo}}\ and\ \bibinfo {author} {\bibfnamefont {Y.}~\bibnamefont {Sumino}},\
  }\Doi {10.1016/j.nuclphysb.2014.10.010} {\bibfield  {journal} {\bibinfo
  {journal} {Nucl. Phys.},\ }\textbf {\bibinfo {volume} {B889}},\ \bibinfo
  {pages} {156} (\bibinfo {year} {2014})},\ \Eprint
  {http://arxiv.org/abs/1408.5590} {arXiv:1408.5590 [hep-ph]} \BibitemShut
  {NoStop}%
\bibitem [{\citenamefont {Chetyrkin}\ \emph {et~al.}(2000)\citenamefont
  {Chetyrkin}, \citenamefont {K{\"u}hn},\ and\ \citenamefont
  {Steinhauser}}]{Chetyrkin:2000yt}%
  \BibitemOpen
  \bibfield  {author} {\bibinfo {author} {\bibfnamefont {K.~G.}\ \bibnamefont
  {Chetyrkin}}, \bibinfo {author} {\bibfnamefont {J.~H.}\ \bibnamefont
  {K{\"u}hn}}, \ and\ \bibinfo {author} {\bibfnamefont {M.}~\bibnamefont
  {Steinhauser}},\ }\Doi {10.1016/S0010-4655(00)00155-7} {\bibfield  {journal}
  {\bibinfo  {journal} {Comput. Phys. Commun.},\ }\textbf {\bibinfo {volume}
  {133}},\ \bibinfo {pages} {43} (\bibinfo {year} {2000})},\ \Eprint
  {http://arxiv.org/abs/hep-ph/0004189} {arXiv:hep-ph/0004189 [hep-ph]}
  \BibitemShut {NoStop}%
\bibitem [{\citenamefont {Pineda}(2003)}]{Pineda:2002se}%
  \BibitemOpen
  \bibfield  {author} {\bibinfo {author} {\bibfnamefont {A.}~\bibnamefont
  {Pineda}},\ }\Doi {10.1088/0954-3899/29/2/313} {\bibfield  {journal}
  {\bibinfo  {journal} {J. Phys.},\ }\textbf {\bibinfo {volume} {G29}},\
  \bibinfo {pages} {371} (\bibinfo {year} {2003})},\ \Eprint
  {http://arxiv.org/abs/hep-ph/0208031} {arXiv:hep-ph/0208031 [hep-ph]}
  \BibitemShut {NoStop}%
\bibitem [{\citenamefont {Bazavov}\ \emph {et~al.}(2014)\citenamefont
  {Bazavov}, \citenamefont {Brambilla}, \citenamefont {Garcia~i Tormo},
  \citenamefont {Petreczky}, \citenamefont {Soto},\ and\ \citenamefont
  {Vairo}}]{Bazavov:2014soa}%
  \BibitemOpen
  \bibfield  {author} {\bibinfo {author} {\bibfnamefont {A.}~\bibnamefont
  {Bazavov}}, \bibinfo {author} {\bibfnamefont {N.}~\bibnamefont {Brambilla}},
  \bibinfo {author} {\bibfnamefont {X.}~\bibnamefont {Garcia~i Tormo}},
  \bibinfo {author} {\bibfnamefont {P.}~\bibnamefont {Petreczky}}, \bibinfo
  {author} {\bibfnamefont {J.}~\bibnamefont {Soto}}, \ and\ \bibinfo {author}
  {\bibfnamefont {A.}~\bibnamefont {Vairo}},\ }\Doi
  {10.1103/PhysRevD.90.074038} {\bibfield  {journal} {\bibinfo  {journal}
  {Phys. Rev.},\ }\textbf {\bibinfo {volume} {D90}},\ \bibinfo {pages} {074038}
  (\bibinfo {year} {2014})},\ \Eprint {http://arxiv.org/abs/1407.8437}
  {arXiv:1407.8437 [hep-ph]} \BibitemShut {NoStop}%
\bibitem [{\citenamefont {Pineda}\ and\ \citenamefont
  {Soto}(2000)}]{Pineda:2000gza}%
  \BibitemOpen
  \bibfield  {author} {\bibinfo {author} {\bibfnamefont {A.}~\bibnamefont
  {Pineda}}\ and\ \bibinfo {author} {\bibfnamefont {J.}~\bibnamefont {Soto}},\
  }\Doi {10.1016/S0370-2693(00)01261-2} {\bibfield  {journal} {\bibinfo
  {journal} {Phys. Lett.},\ }\textbf {\bibinfo {volume} {B495}},\ \bibinfo
  {pages} {323} (\bibinfo {year} {2000})},\ \Eprint
  {http://arxiv.org/abs/hep-ph/0007197} {arXiv:hep-ph/0007197 [hep-ph]}
  \BibitemShut {NoStop}%
\bibitem [{\citenamefont {Brambilla}\ \emph {et~al.}(2009)\citenamefont
  {Brambilla}, \citenamefont {Vairo}, \citenamefont {Garcia~i Tormo},\ and\
  \citenamefont {Soto}}]{Brambilla:2009bi}%
  \BibitemOpen
  \bibfield  {author} {\bibinfo {author} {\bibfnamefont {N.}~\bibnamefont
  {Brambilla}}, \bibinfo {author} {\bibfnamefont {A.}~\bibnamefont {Vairo}},
  \bibinfo {author} {\bibfnamefont {X.}~\bibnamefont {Garcia~i Tormo}}, \ and\
  \bibinfo {author} {\bibfnamefont {J.}~\bibnamefont {Soto}},\ }\Doi
  {10.1103/PhysRevD.80.034016} {\bibfield  {journal} {\bibinfo  {journal}
  {Phys. Rev.},\ }\textbf {\bibinfo {volume} {D80}},\ \bibinfo {pages} {034016}
  (\bibinfo {year} {2009})},\ \Eprint {http://arxiv.org/abs/0906.1390}
  {arXiv:0906.1390 [hep-ph]} \BibitemShut {NoStop}%
\bibitem [{\citenamefont {Pineda}(1998)}]{Pineda:1998id}%
  \BibitemOpen
  \bibfield  {author} {\bibinfo {author} {\bibfnamefont {A.}~\bibnamefont
  {Pineda}},\ }\emph {\bibinfo {title} {{Heavy quarkonium and nonrelativistic
  effective field theories}}},\ \href@noop {} {Ph.D. thesis},\ \bibinfo
  {school} {Barcelona U.} (\bibinfo {year} {1998})\BibitemShut {NoStop}%
\bibitem [{\citenamefont {Hoang}\ \emph {et~al.}(1999)\citenamefont {Hoang},
  \citenamefont {Smith}, \citenamefont {Stelzer},\ and\ \citenamefont
  {Willenbrock}}]{Hoang:1998nz}%
  \BibitemOpen
  \bibfield  {author} {\bibinfo {author} {\bibfnamefont {A.~H.}\ \bibnamefont
  {Hoang}}, \bibinfo {author} {\bibfnamefont {M.~C.}\ \bibnamefont {Smith}},
  \bibinfo {author} {\bibfnamefont {T.}~\bibnamefont {Stelzer}}, \ and\
  \bibinfo {author} {\bibfnamefont {S.}~\bibnamefont {Willenbrock}},\
  }\href@noop {} {\bibfield  {journal} {\bibinfo  {journal} {Phys. Rev.},\
  }\textbf {\bibinfo {volume} {D59}},\ \bibinfo {pages} {114014} (\bibinfo
  {year} {1999})},\ \Eprint {http://arxiv.org/abs/hep-ph/9804227}
  {hep-ph/9804227} \BibitemShut {NoStop}%
\bibitem [{\citenamefont {Beneke}(1998)}]{Beneke:1998rk}%
  \BibitemOpen
  \bibfield  {author} {\bibinfo {author} {\bibfnamefont {M.}~\bibnamefont
  {Beneke}},\ }\Doi {10.1016/S0370-2693(98)00741-2} {\bibfield  {journal}
  {\bibinfo  {journal} {Phys. Lett.},\ }\textbf {\bibinfo {volume} {B434}},\
  \bibinfo {pages} {115} (\bibinfo {year} {1998})},\ \Eprint
  {http://arxiv.org/abs/hep-ph/9804241} {arXiv:hep-ph/9804241 [hep-ph]}
  \BibitemShut {NoStop}%
\bibitem [{\citenamefont {Pineda}\ and\ \citenamefont
  {Signer}(2006)}]{Pineda:2006gx}%
  \BibitemOpen
  \bibfield  {author} {\bibinfo {author} {\bibfnamefont {A.}~\bibnamefont
  {Pineda}}\ and\ \bibinfo {author} {\bibfnamefont {A.}~\bibnamefont
  {Signer}},\ }\Doi {10.1103/PhysRevD.73.111501} {\bibfield  {journal}
  {\bibinfo  {journal} {Phys. Rev.},\ }\textbf {\bibinfo {volume} {D73}},\
  \bibinfo {pages} {111501} (\bibinfo {year} {2006})},\ \Eprint
  {http://arxiv.org/abs/hep-ph/0601185} {arXiv:hep-ph/0601185 [hep-ph]}
  \BibitemShut {NoStop}%
\bibitem [{\citenamefont {Ayala}\ \emph {et~al.}(2014)\citenamefont {Ayala},
  \citenamefont {Cveti\v{c}},\ and\ \citenamefont {Pineda}}]{Ayala:2014yxa}%
  \BibitemOpen
  \bibfield  {author} {\bibinfo {author} {\bibfnamefont {C.}~\bibnamefont
  {Ayala}}, \bibinfo {author} {\bibfnamefont {G.}~\bibnamefont {Cveti\v{c}}}, \
  and\ \bibinfo {author} {\bibfnamefont {A.}~\bibnamefont {Pineda}},\ }\Doi
  {10.1007/JHEP09(2014)045} {\bibfield  {journal} {\bibinfo  {journal} {JHEP},\
  }\textbf {\bibinfo {volume} {09}},\ \bibinfo {pages} {045} (\bibinfo {year}
  {2014})},\ \Eprint {http://arxiv.org/abs/1407.2128} {arXiv:1407.2128
  [hep-ph]} \BibitemShut {NoStop}%
\bibitem [{\citenamefont {Komijani}(2017)}]{Komijani:2017vep}%
  \BibitemOpen
  \bibfield  {author} {\bibinfo {author} {\bibfnamefont {J.}~\bibnamefont
  {Komijani}},\ }\Doi {10.1007/JHEP08(2017)062} {\bibfield  {journal} {\bibinfo
   {journal} {JHEP},\ }\textbf {\bibinfo {volume} {08}},\ \bibinfo {pages}
  {062} (\bibinfo {year} {2017})},\ \Eprint {http://arxiv.org/abs/1701.00347}
  {arXiv:1701.00347 [hep-ph]} \BibitemShut {NoStop}%
\bibitem [{\citenamefont {Ebert}\ \emph {et~al.}(2003)\citenamefont {Ebert},
  \citenamefont {Faustov},\ and\ \citenamefont {Galkin}}]{Ebert:2002pp}%
  \BibitemOpen
  \bibfield  {author} {\bibinfo {author} {\bibfnamefont {D.}~\bibnamefont
  {Ebert}}, \bibinfo {author} {\bibfnamefont {R.~N.}\ \bibnamefont {Faustov}},
  \ and\ \bibinfo {author} {\bibfnamefont {V.~O.}\ \bibnamefont {Galkin}},\
  }\Doi {10.1103/PhysRevD.67.014027} {\bibfield  {journal} {\bibinfo  {journal}
  {Phys. Rev.},\ }\textbf {\bibinfo {volume} {D67}},\ \bibinfo {pages} {014027}
  (\bibinfo {year} {2003})},\ \Eprint {http://arxiv.org/abs/hep-ph/0210381}
  {arXiv:hep-ph/0210381 [hep-ph]} \BibitemShut {NoStop}%
\bibitem [{\citenamefont {Godfrey}\ and\ \citenamefont
  {Moats}(2015)}]{Godfrey:2015dia}%
  \BibitemOpen
  \bibfield  {author} {\bibinfo {author} {\bibfnamefont {S.}~\bibnamefont
  {Godfrey}}\ and\ \bibinfo {author} {\bibfnamefont {K.}~\bibnamefont
  {Moats}},\ }\Doi {10.1103/PhysRevD.92.054034} {\bibfield  {journal} {\bibinfo
   {journal} {Phys. Rev.},\ }\textbf {\bibinfo {volume} {D92}},\ \bibinfo
  {pages} {054034} (\bibinfo {year} {2015})},\ \Eprint
  {http://arxiv.org/abs/1507.00024} {arXiv:1507.00024 [hep-ph]} \BibitemShut
  {NoStop}%
\bibitem [{\citenamefont {Grotch}\ \emph {et~al.}(1984)\citenamefont {Grotch},
  \citenamefont {Owen},\ and\ \citenamefont {Sebastian}}]{Grotch:1984gf}%
  \BibitemOpen
  \bibfield  {author} {\bibinfo {author} {\bibfnamefont {H.}~\bibnamefont
  {Grotch}}, \bibinfo {author} {\bibfnamefont {D.~A.}\ \bibnamefont {Owen}}, \
  and\ \bibinfo {author} {\bibfnamefont {K.~J.}\ \bibnamefont {Sebastian}},\
  }\Doi {10.1103/PhysRevD.30.1924} {\bibfield  {journal} {\bibinfo  {journal}
  {Phys. Rev.},\ }\textbf {\bibinfo {volume} {D30}},\ \bibinfo {pages} {1924}
  (\bibinfo {year} {1984})}\BibitemShut {NoStop}%
\bibitem [{\citenamefont {Shi}(2017)}]{Shi:2016cef}%
  \BibitemOpen
  \bibfield  {author} {\bibinfo {author} {\bibfnamefont {Y.-L.}\ \bibnamefont
  {Shi}},\ }\Doi {10.1140/epjc/s10052-017-4818-5} {\bibfield  {journal}
  {\bibinfo  {journal} {Eur. Phys. J.},\ }\textbf {\bibinfo {volume} {C77}},\
  \bibinfo {pages} {253} (\bibinfo {year} {2017})},\ \Eprint
  {http://arxiv.org/abs/1611.09838} {arXiv:1611.09838 [hep-ph]} \BibitemShut
  {NoStop}%
\bibitem [{\citenamefont {Li}\ and\ \citenamefont {Chao}(2009)}]{Li:2009nr}%
  \BibitemOpen
  \bibfield  {author} {\bibinfo {author} {\bibfnamefont {B.-Q.}\ \bibnamefont
  {Li}}\ and\ \bibinfo {author} {\bibfnamefont {K.-T.}\ \bibnamefont {Chao}},\
  }\Doi {10.1088/0253-6102/52/4/20} {\bibfield  {journal} {\bibinfo  {journal}
  {Commun. Theor. Phys.},\ }\textbf {\bibinfo {volume} {52}},\ \bibinfo {pages}
  {653} (\bibinfo {year} {2009})},\ \Eprint {http://arxiv.org/abs/0909.1369}
  {arXiv:0909.1369 [hep-ph]} \BibitemShut {NoStop}%
\bibitem [{\citenamefont {Abdesselam}\ \emph {et~al.}(2016)\citenamefont
  {Abdesselam} \emph {et~al.}}]{Abdesselam:2016xbr}%
  \BibitemOpen
  \bibfield  {author} {\bibinfo {author} {\bibfnamefont {A.}~\bibnamefont
  {Abdesselam}} \emph {et~al.} (\bibinfo {collaboration} {Belle}),\ }\href@noop
  {} { (\bibinfo {year} {2016})},\ \Eprint {http://arxiv.org/abs/1606.01276}
  {arXiv:1606.01276 [hep-ex]} \BibitemShut {NoStop}%
\bibitem [{\citenamefont {Segovia}\ \emph {et~al.}(2012)\citenamefont
  {Segovia}, \citenamefont {Entem}, \citenamefont {Fernandez},\ and\
  \citenamefont {Ruiz~Arriola}}]{Segovia:2011tb}%
  \BibitemOpen
  \bibfield  {author} {\bibinfo {author} {\bibfnamefont {J.}~\bibnamefont
  {Segovia}}, \bibinfo {author} {\bibfnamefont {D.~R.}\ \bibnamefont {Entem}},
  \bibinfo {author} {\bibfnamefont {F.}~\bibnamefont {Fernandez}}, \ and\
  \bibinfo {author} {\bibfnamefont {E.}~\bibnamefont {Ruiz~Arriola}},\ }\Doi
  {10.1103/PhysRevD.85.074001} {\bibfield  {journal} {\bibinfo  {journal}
  {Phys. Rev.},\ }\textbf {\bibinfo {volume} {D85}},\ \bibinfo {pages} {074001}
  (\bibinfo {year} {2012})},\ \Eprint {http://arxiv.org/abs/1108.0208}
  {arXiv:1108.0208 [hep-ph]} \BibitemShut {NoStop}%
\end{thebibliography}%

\end{document}